\documentclass[12pt]{article}
\usepackage{a4wide}
\usepackage[centertags]{amsmath}
\usepackage{amssymb}
\usepackage[sort&compress,numbers]{natbib}
\usepackage{ifpdf}
\usepackage{setspace}
\usepackage{amsfonts}
\usepackage{color}
\usepackage{threeparttable}
\usepackage{cancel}
\ifpdf
\usepackage[colorlinks=true,
linkcolor=black,
citecolor=black,
urlcolor=blue,
filecolor=blue,
pdfstartview=FitV,
pdftitle={},
pdfauthor={},
pdfsubject={Hydrodynamics},
pdfkeywords={hydrodynamics, AdS/CFT},
pdfpagemode=None,
bookmarksopen=true
]{hyperref}

\else
\usepackage[hypertex]{hyperref}
\fi
\usepackage{rotating}
\makeatletter
\@addtoreset{equation}{section}

\makeatletter
\renewcommand\section{\@startsection {section}{1}{\z@}%
	{-3.5ex \@plus -1ex \@minus -.2ex}
	{2.3ex \@plus.2ex}%
	{\normalfont\large\bfseries}}
\renewcommand\subsection{\@startsection{subsection}{2}{\z@}%
	{-3.25ex\@plus -1ex \@minus -.2ex}%
	{1.5ex \@plus .2ex}%
	{\normalfont\bfseries}}

\def\sec#1{\S \;\ref{#1}}

\title{{ Magneto-transport in an anomalous fluid with weakly broken symmetries, in weak and  strong regime}}

\author{Navid Abbasi$^{a}$\footnote{abbasi@ipm.ir}, \ Armin Ghazi$^{b}$\footnote{armin.ghazi@physics.sharif.edu},  \ Farid Taghinavaz$^{a}$\footnote{ftaghinavaz@ipm.ir}, \ Omid Tavako$l^{b}$\footnote{omid.t721@gamil.com}\\
	\small{\emph{$^{a}$  School of Particles and Accelerators, Institute for Research in Fundamental Sciences (IPM)}}, \\
	\small{\emph{P.O. Box 19395-5531, Tehran, Iran}}\\
	\small{\emph{$^{b}$Department of Physics, Sharif University of Technology,}} \\
	\small{\emph{P.O. Box 11365-9161, Tehran, Iran}} \\ [1mm]
}

\begin{document}
	
	\setlength{\baselineskip}{16pt}
	\begin{titlepage}
		\maketitle
		
		\vspace{-36pt}
		
		\begin{abstract}
			We consider a fluid with  weakly broken time and translation symmetries. We assume the fluid also possesses a $U(1)$ symmetry which is not only weakly broken, but is anomalous.  
			We use the second order chiral quasi-hydrodynamics to compute the magneto conductivities of this fluid in the presence of a weak magnetic field. Analogous to the electrical and thermoelectric conductivities, it turns out that the thermal conductivity depends on the coefficient of mixed gauge-gravitational anomaly. Our results can be applied to the hydrodynamic regime of every arbitrary system, once the thermodynamics of that system is known. By applying them to a free system of Weyl fermions at low temperature limit $T\ll \mu$, we find that our fluid is Onsager reciprocal if the relaxation in all energy, momentum and charge channels occurs at the same rate. In the high temperature limit $T\gg \mu$, we consider a strongly coupled $SU(N_c)$ gauge theory with $N_c\gg1$.  Its holographic dual in thermal equilibrium is a magnetized charged brane from  which, we compute the thermodynamic quantities  and subsequently evaluate the conductivities in gauge theory.
			On the way, we show that analogous to the weak regime in the system of Weyl fermions, an energy cutoff emerges to regulate the thermodynamic quantities in the strong regime of boundary gauge theory.  From this gravity background we also find the coefficients of chiral magnetic effect in agreement with the well-known result of Son-Surowka.

		\end{abstract}
		\thispagestyle{empty}
		\setcounter{page}{0}
	\end{titlepage}

	\renewcommand{\baselinestretch}{1}  
	\tableofcontents
	\renewcommand{\baselinestretch}{1.2}  
	\section{Introduction}
Chiral fluid is a system wherein the microscopic quantum field theory anomalies may be realized macroscopically. The induction of current in the direction of the magnetic field, namely the chiral magnetic effect, is a well-known example in this regard \cite{Fukushima:2008xe}. In the hydrodynamic limit, the triangle anomaly manifests itself through the non-dissipative chiral transport \cite{Son:2009tf} (see the review \cite{Landsteiner:2016led} and references therein). The coefficient of chiral anomaly has been shown to appear in the anomalous transport coefficients, both in weak \cite{Landsteiner:2012kd} and strong coupling \cite{Banerjee:2008th,Erdmenger:2008rm} regimes \footnote{Anomalous transport and its related interesting effects have been widely studied in the literature \cite{Landsteiner:2011cp,Landsteiner:2011iq,Abbasi:2015saa,Yamamoto:2015ria,Chernodub:2015gxa,Chernodub:2018era,Kharzeev:2010gd}.}.  Interestingly, both free quantum field theory and  the gauge/gravity computations reveal that 
 the mixed gauge-gravitational anomaly coefficient contributes to the anomalous  transport, too \cite{Landsteiner:2012kd}. In \cite{Jensen:2012kj}, it was shown that the latter would happen in the hydrodynamics via a jump in the derivative expansion;  the mixed gauge-gravitational anomaly coefficient appears in the coefficients at two orders of derivatives lower than what expected from the hydrodynamic equations of motion.

Chiral transport has also attracted interest in the condensed matter both theoretically and experimentally. Based on Nielsen-Ninomiya arguments on Weyl semimetals (WSM) \cite{Nilson}, it has been shown that the chiral anomaly causes an increase in the longitudinal electrical DC conductivity in a WSM with a quadratic dependence on the weak magnetic field. This is also sometimes referred to as the negative magneto-resistivity \cite{Weylsemimeta:Son}. In the strong magnetic field when a 3-dimensional metal is effectively one dimensional, the conductivity grows linearly in $B$, due to the spectral flow of the lowest Landau level \cite{Gorbar:2013dha}.

In the experiment, an observation of the chiral magnetic effect was reported through the measurement of a large negative magneto-resistance
in zirconium pentatelluride, ZrTe5 \cite{Li:2017mbd}.  Analogously, the observation of a quadratic dependence of thermoelectric DC conductivity on the magnetic field in NbP WSM was recently realized as a signature of the mixed gauge-gravitational anomaly  \cite{Gooth:2017mbd}. The theoretical model based on which the latter was concluded, assumes that the metal is a system with two weakly broken symmetries; time translation together with an anomalous $U(1)$ charge. A weakly broken symmetry may be characterized with a large relaxation time parameter. In \cite{Gooth:2017mbd} and in the steady state, the electrical and thermo-electric conductivities were shown to linearly depend on the relaxation time parameter characterizing the time scale of the inter-valley scattering. 

Independent of WSM physics, in the current paper we extend the computations of \cite{Gooth:2017mbd} to a more general chiral system in which, in addition to energy and charge, the translation is considered as a weakly broken symmetry as well. 
The electrical DC conductivity in such system was already studied in the hydrodynamic regime and also holographically in the strong coupling limit, in \cite{Landsteiner:2014vua}. We first argue that in order to obtain the quadratic dependence of the conductivities on the magnetic field in the hydrodynamic regime, two conditions have to be fulfilled; first the relaxation time parameters $\tau_c$, $\tau_e$ and $\tau_m$, corresponding to the relaxation of charge, energy and momentum,  must be all of the order of inverse of the gradients in the system. This ensures that the system is described with the theory of quasi-hydrodynamics recently developed in  \cite{Grozdanov:2018fic} (and also an earlier related work 		\cite{Grozdanov:2017kyl}). Secondly, let us recall that in the standard hydrodynamic computations, the background magnetic field is assumed to be as a one-derivative data \cite{Son:2009tf}. By the motivation coming from the WSMs in which the magneto conductivities quadratically depend on the background magnetic field \cite{Weylsemimeta:Son}, we are to use the second order anomalous hydrodynamics which naturally admits terms quadratic in the magnetic filed. 

Then by turning on a weak  electric field and a small thermal gradient in the system, we compute the longitudinal conductivities in the system via studying the linear response of the system, namely by reading the electric and heat currents, $J_i$ and $Q_i$
\begin{equation}\label{magnet_conductivity}
\begin{pmatrix}
J_i  \\
Q_i\end{pmatrix}=\begin{pmatrix}
\sigma_{} &T \alpha_{1} \\
T \alpha_{2} &T \kappa_{} 
\end{pmatrix}\,\begin{pmatrix}
E_j  \\
-\nabla_j T/T\end{pmatrix}.
\end{equation}              
In addition to previously-reported properties of the magneto-electrical conductivity in \cite{Landsteiner:2014vua}, we find the following two
new aspects of conductivities; first it turns out that just like the electrical conductivity, the other conductivities quadratically depend on the magnetic field in the common form
\begin{equation} 
\sim\, B^2\,\left(c^2\mathcal{A}+c\, c_g\, \mathcal{B}+c_g^2\,\mathcal{C}\right)
\end{equation}
where $c$ and $c_g$ are the coefficients of chiral and mixed gauge-gravitational anomalies, respectively. The coefficients $\mathcal{A}$, $\mathcal{B}$ and $\mathcal{C}$ contain the thermodynamic information and are non-vanishing in general. The second point with the conductivities in \eqref{magnet_conductivity}, which is more important, is that they are not Onsager reciprocal. As it is well-known, to the only process which contributes to the non-magnetic part of the conductivities is the momentum relaxation and from it, the Onsager reciprocal symmetry is immediately concluded \cite{Lucas:2017idv}. We show that the inequality of the relaxation rate for charge and energy with respect to that of momentum (as well as to each other) spoils the reciprocal symmetry in the magneto-transport. However as we argue in the following, there is still a situation in which the symmetry is restored. 

Let us recall that the hydrodynamic results are general in the sense that no any specific thermodynamic equation of state is needed to obtain them. We use our hydro results to find the magneto-conductivities in two opposite certain limits. First we apply them to a free system of Weyl fermions in the presence of magnetic field. Such system has been extensively studied in the context of chiral kinetic theory in recent years \cite{Son:2012wh,Gao:2012ix,Son:2012zy,Hidaka:2016yjf,Chen:2015gta,Chen:2014cla,Stephanov:2012ki,Abbasi:2017tea}. As was mentioned in \cite{Abbasi:2018zoc}, to study the magneto-transport in a relativistic Weyl fluid it is needed to take into account the quantum corrections to second order. The magneto-conductivities as well as the effect of the second order corrections on the thermodynamics of the system were computed in the mentioned paper. We take the thermodynamic information from  \cite{Abbasi:2018zoc} as the only input needed to evaluate our hydrodynamic results in the systems of free Weyl particles. After finding the conductivities for such system, we compare them with those directly obtained from chiral kinetic theory in  \cite{Abbasi:2018zoc}. It turns out that in the limit $\tau_{c}=\tau_e=\tau_m$, the microscopic time reversal symmetry emerges as reciprocal symmetry in the kinetic coefficients of the conductivity. Surprisingly at the same limit, the hydro conductivities obey the constraints obtained from Ward identity in \cite{Abbasi:2018zoc}. This may suggest that in this special limit, the quasi-hydrodynamic equations, namely the hydro equations with relaxation time terms, can be derived from a generating functional with modified gauge and diffeomorphism symmetries \cite{Abbasi:2019zoc}.    

 As the second case, we consider a strongly coupled $\mathcal{N}=4$ super Yang-Mills $SU(N_c)$ gauge theory at the limit $N_c\gg1$. As in the case of Weyl fluid, to study the quasi-hydrodynamic magneto-transport in the system we only need to know the thermodynamics of the system.
 Using gauge/gravity duality \cite{Maldacena:1997re,Witten:1998qj} we 
 compute the thermodynamic quantities of the system from the dual gravity solution.
 Such solution  may be described by a weakly magnetized brane in the AdS5 \cite{DHoker:2009ixq}.
 We use the general non-evaluated solution of the mentioned paper in the limit $B\ll T^2$ and analytically evaluate it in the limit  $B\ll\mu^2\ll T^2$. Having found the solution in a double expansion over $B$ and $\mu/T$, we then compute the thermodynamic quantities associate with the boundary theory. Using them, we can evaluate the hydro results in this system and compute the magneto-conductivities as well. 
 
 An interesting point with our holographic computations is the emergence of an energy scale in the boundary theory. This is actually originated from the appearance of a term $\ln \frac{\pi T}{L}$, with $L$ being the radius of AdS5, in the boundary quantities. The fact that physical quantities on the boundary have to be given just in terms of boundary data, forces to remove $L$ from all physical expressions. We will show that by adding a finite appropriate counter term to the boundary stress tensor, $L$ can be replaced with a boundary energy scale, say $\Delta$. A similar behavior in a chiral system in the presence of magnetic field was already observed in \cite{Abbasi:2018zoc}. In the mentioned paper, the emergent energy scale was an IR cutoff for the momentum of Weyl particles in the momentum space.  
 
 The rest of the papers is organized as it follows.
 In \sec{hydro}, we first give the quasi-hydrodynamic equations to second order in derivatives and then compute the magneto-conductivities via studying the linear response of the system to the external sources. In \sec{comparison_hydro_CKT} we apply our hydrodynamic results to the system of free Weyl fermions. Onsager reciprocal symmetry is also studied in this section. 
In \sec{2}, we holographically compute the magneto-conductivities in the strong limit. We end in \sec{conclusion} with giving a brief review of results and discussing on follow up questions.

\section{Anomalous Transport from Hydrodynamics}
\label{hydro}

What we are going to do in this section is to investigate how  dissipative and anomalous hydrodynamic transport may contribute to the magneto-transport in a Weyl fluid.  The second order anomalous hydrodynamic constitutive relations and the constraints on the corresponding transport coefficients has been developed in \cite{Kharzeev:2011ds}\footnote{We rewrite the constitutive relations of  \cite{Kharzeev:2011ds} for the "thermodynamic" frame\footnote{Somewhere in the literature, it is referred to as the Laboratory frame as well.}, in which the rest frame of the fluid elements is nothing but the laboratory frame, to all orders in the derivative expansion. Since the thermodynamic frame is the hydrodynamic frame consistent with the chiral kinetic theory, we can then easily compare the results of the current section with those obtained earlier from chiral kinetic theory. }. Considering \cite{Kharzeev:2011ds} and using the linear response method, we compute the response of the fluid to turning the background electric field as well as the background gradient of temperature on.

\subsection{Second order hydrodynamics in  an anomalous system }
In a simple system with just one $U(1)$ anomalous charge, the hydrodynamic equations are the energy-momentum conservation together with the non-conservation equation of the charge due to anomaly. One writes:
\begin{eqnarray}\label{EoMT}
\partial_{\mu}T^{\mu \nu}= F^{\nu \alpha} J_{\alpha}\\\label{EoMJ}
\partial_{\mu}J^{\mu}=\,c \,E^{\mu}B_{\mu}
\end{eqnarray}
with $c$ being the coefficient of chiral anomaly and $E^{\mu}$ and $B^{\mu}$ the covariant background electric and magnetic field. The corresponding constitutive relations for stress tensor and charge current are
\begin{equation}
\label{Tmunu_general}
\begin{split}
T^{\mu \nu}=& \,(\epsilon+p) u^{\mu} u^{\nu}+ p \,\eta^{\mu \nu} +\tau^{\mu \nu},\\
J^{\mu}=& \,n\, u^{\mu} +\nu^{\mu}.
\end{split}
\end{equation}
where the so-called derivative corrections $\tau^{\mu \nu}$ and $\nu^{\mu}$ can be specified in terms of the gradient of hydrodynamic fields as well as the slowly varying background metric and gauge fields. Due to reasons mentioned earlier, we have to consider the corrections up to second order in gradients. However, since we are going to compute the conductivities at the limit $k \rightarrow 0$, the dissipative terms can be dropped. We would also rather to work in the Laboratory frame\footnote{see \cite{Abbasi:2017tea} for discussion on the difference between the Laboratory frame and the Landau-Lifshitz one.}. We may write \cite{Kharzeev:2011ds,Hernandez:2017mch}
\begin{eqnarray}
\label{tau_mu_nu}
\tau^{\mu \nu} &=&\sigma^{\mathcal
	\epsilon}_{\mathcal{B}}(u^{\mu}B^{\nu}+u^{\nu}B^{\mu})+ C_1(B^{\mu}B^{\nu}-\frac{1}{3}P^{\mu \nu}B^2)+\Pi^{\mu \nu}_{\alpha \beta}B^{\alpha}\big(C_2 E^{\beta}-C_3 \mu\frac{\nabla^{\beta} T }{T} \big)\label{}\\
\nu^{\mu} & = &\sigma_{E}\, \left(E^{\mu}-T P^{\mu \nu}\nabla_{\nu}\frac{\mu}{T}\right)+\sigma_{\mathcal{B}}\, B^{\mu}\label{nu_mu}\\
\Pi^{\mu \nu}_{\alpha \beta}&=&\frac{1}{2}(P^{\mu}_{\,\,\, \alpha}P^{\nu}_{\,\,\, \beta}+P^{\mu}_{\,\,\, \beta}P^{\nu}_{\,\,\, \alpha}-\frac{2}{3}P^{\mu\nu}P_{\alpha \beta})
\end{eqnarray}
where $P^{\mu \nu}= u^{\mu}u^{\nu}+\eta^{\mu \nu}$ and $C_1$, $C_2$ and $C_3$ are the unknown coefficients, being functions of the thermodynamics of the matter. The anomalous transport coefficients are given in terms of the thermodynamic variables as well as the anomaly coefficients \cite{Landsteiner:2012kd}: 
\begin{eqnarray}\label{}
\sigma^{\mathcal
	\epsilon}_{\mathcal{B}}=\frac{1}{2}\big(c \mu^2+c_g T^2\big),\,\,\,\,\,
\sigma_{\mathcal{B}}=c\, \mu
\end{eqnarray}
with $c$ and $c_g$ being the coefficients of chiral anomaly  and the mixed gauge-gravitational anomaly.

One may criticize that in contrary to our previous emphasis on dropping the dissipative terms in the $k\rightarrow 0$ limit, we have kept two of such terms in the constitutive relations; the term including $\nabla T$ in $\tau^{\mu \nu}$ together with  $\sigma_{E} E_{\mu}$ in $\nu^{\mu}$. In fact, we get $\boldsymbol{\nabla}T$ and $\boldsymbol{E}$ as the external sources which inject energy into the system and so continue to keep them turned on even in the uniform state of the system with $k=0$. Let us also note that the conductivity we are searching for is not $\sigma_{E}$ in equation \eqref{nu_mu}; the latter is the contribution which depends on the microscopic of the system and cannot be found from hydrodynamics. Following \cite{Hartnoll:2007ih}, we call it the quantum conductivity in this note.
In the following we will explain that $\sigma_{E}$ is only one of the several parts contributing to the total conductivity.

\subsection{Linear response of the system to the external sources in the framework of quasi-hydrodynamics}
Our strategy to find the conductivities is to turn on the external sources
  $\boldsymbol{\nabla}T$ and $\boldsymbol{E}$ and then study the linear response of the system to the sources, in the hydrodynamic regime and to second order in derivative expansion. Although according to the standard derivative counting of the hydrodynamics,  $\boldsymbol{\nabla}T$ and $\boldsymbol{E}$ are basically the one-derivative terms, we never consider terms including the multiplication of these terms either with themselves or with each other; this is just for the consistency with the linear response regime in which, one has to take the sources to linear order.
    
  In order to take into account the anomalous effects, we turn on a weak background magnetic field, $B\ll T^2$, in the $z$ direction.
  The longitudinal conductivities then  may be computed by taking the external sources to be directed in the $z$ direction as well.
  
  Before turning on the sources and just by considering $F^{xy}=-F^{yx}=B$, the only non-vanishing stress tensor and charge current components are
   \begin{eqnarray}
   \tilde{T}^{00}&=&\epsilon,\,\,\,\,\,\,\,\,\,\,\,\,  \tilde{T}^{zz}=p,\,\,\,\,\,\,\,\,\,\,\,\,\,\,\, \tilde{T}^{ii}= p -\frac{C_1 }{3}B^{2}\,\, (i=x,y),\,\,\,\,\,\,\,\,\,  \tilde{J}^{0}=n\\
   \tilde{T}^{0z}&=&\frac{1}{2}\big(c_g T^2+c \mu^2\big)B,\,\,\,\,\,\,\,\,\,\,\,\,\, \tilde{J}^{z}= c \mu \,B
   \end{eqnarray}
 where the tilde over the components denotes that they are being computed in the equilibrium. When perturbing the equilibrium of the system by the weak external sources $\delta \boldsymbol{E}$ and $\delta\boldsymbol{\zeta}=-\boldsymbol{\nabla}\log T$, which are actually the gradient of some long wave-length background fields, the hydro fields would respond as (See Appendix \ref{Sources} for  details of  the source terms.)
   \begin{eqnarray}
   u^{\mu}&\rightarrow&\big(1,\delta u_x(\boldsymbol{x},t),\delta u_y(\boldsymbol{x},t),\delta u_z(\boldsymbol{x},t)\big)\\
   T&\rightarrow&T(\boldsymbol{x})+ \delta T(\boldsymbol{x},t)\\
   \mu&\rightarrow&\mu+ \delta \mu(\boldsymbol{x},t)
   \end{eqnarray}
   and following them, so do the thermodynamic variables, according to the equation of state of the fluid in the equilibrium. One writes
  \begin{eqnarray}\label{epsilon_thermo}
  \epsilon&\rightarrow&\epsilon+ e_1\, \delta \mu(\boldsymbol{x},t)+ e_2\, \delta T(\boldsymbol{x},t)=\,\epsilon+ \big(\frac{\partial \epsilon}{\partial \mu}\big)\biggr\rvert_T\delta \mu+ \big(\frac{\partial \epsilon}{\partial T}\big)\biggr\rvert_{\mu} \delta T\\
  p&\rightarrow&p+ n\, \delta \mu(\boldsymbol{x},t)+ s\, \delta T(\boldsymbol{x},t)\\ \label{n_thermo}
  n&\rightarrow&n+ f_1\, \delta \mu(\boldsymbol{x},t)+ f_2\, \delta T(\boldsymbol{x},t)=\,n+ \big(\frac{\partial n}{\partial \mu}\big)\biggr\rvert_T\delta \mu+ \big(\frac{\partial n}{\partial T}\big)\biggr\rvert_{\mu} \delta T.
  \end{eqnarray}
It is then easy to show that the $U(1)$ field strength is perturbed as well.
While in the absence of the sources the only external field is assumed to be a constant magnetic field, by turning on the  electric field, fluid flows and the background magnetic field is seen as an  electric field in the rest frame of the moving fluid elements. In the laboratory frame and by considering the source electric field in the $z$-direction, e.g. $\delta E_z$,  we may write
\begin{equation*}\label{MatrixB}
	\tilde{F}^{\mu\nu}+\delta F^{\mu\nu}=\begin{pmatrix}
		0 & -\delta u_y B &-\delta u_x B & \delta E_z \\
		\delta u_y B &  0 &B& 0 \\
		\delta u_x B  & -B &0&0\\
		-\delta E_z & 0 &0&0
	\end{pmatrix}.
\end{equation*}
From now on for brevity, we omit the arguments of the fields.
In order to find the linear equations governing the out of equilibrium response of the hydrodynamic fields to the sources, it is needed to  linearize the energy momentum tensor and charge current as well. The stress tensor components are perturbed as 
\begin{align}\label{deltaTtz}
\delta T^{0i}&=w \delta u_i,\,\,\,\,\,\,\,(i=x,y)\\
\delta T^{0z}&=w \delta u_z+\big(c_g T\, \delta T+c \mu\, \delta \mu\big)B\\
	\delta T^{00}&=\,\big(c_g T^2+c \mu^2\big)\delta u_z \,B+e_1 \delta \mu+\,e_2 \delta T\\\label{deltaT_ii}
	\delta T^{ii}&=\,s \,\delta T+\,n\, \delta \mu -\,\frac{ C_2 }{3} B\,\delta E_z -\frac{ C_3}{3 }\,B\,\frac{\mu}{T}\delta\zeta_z,\,\,\,\,(i=x,y)\\
	\delta T^{zz}&=\,\big(c_g T^2+c \mu^2\big) B\,\delta u_z +n \delta \mu+\,s \delta T+\frac{ 2C_2 }{3} B\,\delta E_z +\frac{ C_3}{3 }\,B\,\frac{\mu}{T}\delta\zeta_z\\
	\delta T^{zi}&=\frac{1}{2}\big(c_g T^2+c \mu^2\big) B\,\delta u_i ,\,\,\,\,\,\,\,\,(i=x,y).
\end{align}
It has to be noted that the perturbations of the velocity in the transverse directions, namely $\delta u_x$ and $\delta u_y$, do not appear in the components written above. They only enter through $T^{0i}$ and $T^{zi}$ ($i=x,y$) and these components clearly do not contribute to the longitudinal transport (This point becomes clear in the following.).
The linear perturbation of the charge current components are given by \footnote{Let us denote that according to \eqref{magnet_conductivity}, in order to compute the conductivities we need the two-point retarded Green's functions. Consequently we only have to study the response of the system in the linear order. The response of the system at quadratic order in sources which leads to finding the 3-point retarded Green's functions is studied when deriving the Kubo formulas for the second order transport coefficients 	\cite{Moore:2012tc}.}
\begin{eqnarray}
\delta J^{0}&=&c \mu\, \delta u_z B+f_1 \,\delta \mu+\,f_2 \,\delta T\\
\delta J^{x}&=&n\, \delta u_x+2 \sigma_E \delta u_y B\\
\delta J^{y}&=&n\, \delta u_y-2 \sigma_E \delta u_x B\\\label{delta_j_z}
\delta J^{z}&=&\,n\delta u_z+\, \sigma_E \delta E_z-\,\sigma_{E}\,\mu\,\delta\zeta_z +\, c  B\,\delta \mu.
\end{eqnarray}
We are now at the point that can study the dynamics of linear perturbations in the hydrodynamic regime. As emphasized earlier, in the $k\rightarrow0$ limit we neglect the momentum diffusion in the fluid and correspondingly drop the terms containing the gradient of hydrodynamic fields. However, even in a spatially uniform state, there still may exist another dissipative process in the fluid. The energy pumped into the system by the sources may be dissipated via a scattering mechanism. One can effectively describe such dissipation by simply adding a relaxation time term in the conservation equations. We do so and allow the linearized hydrodynamic equations to get such modification:
\begin{eqnarray}\label{E1}
\partial_{\mu} \delta T^{\mu 0}&=&\delta(F^{0\nu}J_{\nu})-\frac{\delta T^{0\mu}}{\tau_e}\tilde{u}_{\mu}\\\label{E2}
\partial_{\mu} \delta T^{\mu i}&=&\delta(F^{i\,\nu}J_{\nu})-\frac{\delta T^{i\mu}}{\tau_m}\tilde{u}_{\mu}\\\label{E3}
\partial_{\mu} \delta J^{\mu}&=&c\,\delta E_{\mu}  B^{\mu}-\frac{\delta J^{\mu}}{\tau_c}\tilde{u}_{\mu}
\end{eqnarray}
where as before in this section, $\delta$ denotes the linear perturbation around equilibrium state. The above equations are the so-called linear quasi-hydrodynamics equations. Following \cite{Landsteiner:2014vua}, we first assume different hydrodynamic fields relax towards equilibrium with different relaxation time parameters, $\tau_{E}$, $\tau_{M}$ and $\tau_{C}$ for the energy, momentum and anomalous charge, respectively. 

Since we would like to study the response of the system to the sources turned on instantaneously in the system,  we make a Laplace transform in the direction of time in the above three linearized equations\footnote{Considering $\delta\phi$ as each of the fields $\delta \mu$, $\delta T$, $\delta u_z$ and $\delta E_z$, we have
\begin{equation*}
\mathcal{L}[\delta \phi,\omega]=\lim_{\epsilon\rightarrow 0^{+}}\int_{0}^{+\infty}e^{-i \omega t-i \epsilon t}\delta \phi(\boldsymbol{x},t)dt \equiv\, \delta\hat{\phi}(\boldsymbol{x})\,\,\,\rightarrow\,\,\,\,\,\,\mathcal{L}[\partial_t\delta \phi,\omega]=\,-i\omega \delta\hat{\phi}(\boldsymbol{x})- \delta \phi^{(0)}(\boldsymbol{x}) 
\end{equation*}
where $\delta \phi^{(0)}$ denotes the value of the field $\delta \phi$ just at the instant that the sources are turned on.} and following that, make a Fourier transform in the spatial directions. When $k\rightarrow 0$, we find
\begin{align}\label{Laplac_E}
	\left(-i \omega+ \frac{1}{\tau_e}\right)\left(e_1 \delta \hat{\mu} +e_2 \delta \hat{T}+\frac{}{}\big(c_g T^2+c \mu^2\big) B\,\delta \hat{u}_z \right)&=c \mu\, B\, \delta \hat{E}_z^{(0)} +c_g T^2\,  B\,\delta\hat{\zeta}^{(0)}+\cdots \\ \label{Laplac_M}
	\left(-i \omega+ \frac{1}{\tau_m}\right)\left(\big(c_g T \,\delta \hat{T}+c \mu \,\delta \hat{\mu}\big) B+\frac{}{}w \delta \hat{u}_z\right)&= n\, \delta \hat{E}_z^{(0)} +s T\, \delta\hat{\zeta}_z ^{(0)}+\cdots \\ \label{Laplac_C}
	\left(-i \omega+ \frac{1}{\tau_c}\right)\left(f_1 \delta \hat{\mu} +f_2 \delta \hat{T}+c \mu B \frac{}{}\delta \hat{u}_z \right)&=c \,  B\,\delta \hat{E}_z^{(0)}+\cdots.
\end{align}
The dots in these equations (and also below) include terms which depends linearly on either $\delta \mu^{(0)}$ or $\delta T^{(0)}$ or $\delta u_z^{(0)}$, however,  by stetting the initial value of the hydro fields to zero, such terms will no longer contribute.

For further simplifications, we  define
\begin{equation}
\omega_{e}=\,\omega +\frac{i}{\tau_{e}},\,\,\,\,\omega_{m}=\,\omega +\frac{i}{\tau_{m}},\,\,\,\,\,\omega_{c}=\,\omega +\frac{i}{\tau_{c}},
\end{equation}
From the set of Algebraic equations \eqref{Laplac_E}, \eqref{Laplac_M} and \eqref{Laplac_C}, one simply finds the response of the hydrodynamic fields as it follows. In Appendix \ref{hydro_general_result} we give the results for a general case, however, for sake of brevity, here, we just write down the expressions for the simple case $\tau_{c}=\tau_{m}=\tau_{e}$:
\begin{align}\label{delta_T}
\delta \hat{T}&=-\frac{c_g f_1 T+ c (e_1- f_1 \mu)s}{w_0(e_2 f_1 -e_1 f_2)}\,T(\delta \hat{E}_z^{(0)} + \mu \, \delta\hat{\zeta}_z^{(0)})\,\frac{i B}{\omega+\frac{i}{\tau}}+\cdots\\\label{delta_mu}
\delta \hat{\mu}&=\frac{c_g f_2 T+ c (e_2- f_2 \mu)s}{w(e_2 f_1 -e_1 f_2)}\,T(\delta \hat{E}_z^{(0)} + \mu \, \delta\hat{\zeta}_z^{(0)})\,\frac{i B}{\omega+\frac{i}{\tau}}+\cdots\\ \label{delta_u}
\delta \hat{u}_z&=\left(\frac{n}{w}+\frac{c_g^2 f_1 n T^2-c^2(e_2-f_2 \mu)\mu s+ c c_g T \big((e_1- f_1 \mu)s+f_2  \mu n\big)}{w^2(e_2 f_1 -e_1 f_2)}\,T \,B^2\right)\frac{i\, \delta \hat{E}_z^{(0)}}{\omega+\frac{i}{\tau}} \\\nonumber
&-\left(-\frac{s}{w}+\frac{c_g^2 f_1 n T^2-c^2(e_2-f_2 \mu_0)\mu s+ c c_g T \big((e_1- f_1 \mu)s+f_2  \mu n\big)}{w^2(e_2 f_1 -e_1 f_2)}\mu \,B^2\right) \frac{i T\,\delta\hat{\zeta}_z^{(0)} }{\omega+\frac{i}{\tau}}+\cdots.
\end{align}
As before the dots include terms which just linearly depend on the initial value of the hydrodynamic fields. 
Having found the response of the hydrodynamic fields, we can now simply substitute them into the charge and heat current  to read the conductivities.
\subsection{Magneto-transport from hydrodynamics}
\label{magneto_transport_from_hydro}
Due to the presence of a finite chemical potential, the finite charge density mixes the heat and electric currents and consequently, the Ohm's law must be generalized to
\begin{equation}\label{generalized_Ohm}
\begin{pmatrix}
J_i  \\
Q_i\end{pmatrix}=\begin{pmatrix}
\sigma_{ij} &T \alpha_{ij} \\
T \alpha_{ij} &T \kappa_{ij} 
\end{pmatrix}\,\begin{pmatrix}
\delta E_j  \\
\delta\zeta_j\end{pmatrix}.
\end{equation}              
where $J^{i}$ and $Q^i$ are electric and heat currents, respectively.

The longitudinal current induced due to linear response of the system with respect to the external sources has been given in \eqref{delta_j_z}. By substituting the solutions found from the hydrodynamic equations, namely \eqref{delta_T}, \eqref{delta_mu} and \eqref{delta_u}, the induced current \eqref{delta_j_z} takes the form of the \eqref{generalized_Ohm}. So the electric  and the thermoelectric conductivity coefficients simply read\footnote{In \cite{Landsteiner:2014vua}, only the electric conductivity has been computed in the same framework. They have performed the computations in the Landau-Lifshitz frame.}
\begin{align}\label{sigma}
\sigma\,=&\,\sigma_E\,+\frac{i}{\omega_m}\frac{n^{2}}{w}\\\nonumber
&\,\,\,\,\,\,\,\,\,\,\,\,\,+\frac{i B^2\,T}{w^2(e_2f_1-f_2e_1)}\bigg\{c_g^2\frac{f_1T^2n}{\omega_m}+c^2\mu s
\left(e_2(\frac{w}{\omega_C}-\frac{\mu\,n}{\omega_m})-f_2(\frac{w}{\omega_E}-\frac{\mu\,n}{\omega_m}\big)\right)\\&\nonumber
\,\,\,\,\,\,\,\,\,\,\,\,\,\,\,\,\,\,\,\,\,\,\,\,\,\,\,\,\,\,\,\,\,\,\,\,\,\,\,\,\,\,\,\,\,\,\,\,\,\,\,\,\,\,\,\,\,\,\,\,\,\,\,\,+cc_g n\left(e_1(\frac{w}{\omega_C}-\frac{\mu\,n}{\omega_m})-f_1(\frac{w}{\omega_E}-\frac{\mu\,n}{\omega_m})\mu+\frac{f_2T^2s}{\omega_m}\right)\bigg\}
\end{align}
\begin{align}\label{alpha1}
\alpha_1=&-\mu\,\sigma_E+\frac{i}{\omega_m}\frac{Tn  s}{w}\\\nonumber&
+\frac{i B^2}{(e_2f_1-e_1f_2)w^2}\biggl\{c^2\frac{\,\mu s^2 T^2}{\omega_m}(\mu f_2-e_2)
-c_g^2\,\,T^3 n \big(w(\frac{1}{\omega_e}-\frac{1}{\omega_m})+\frac{\mu\,n}{\omega_m}\big)f_1\\&\nonumber \,\,\,\,\,\,\,\,\,\,\,\,\,\,\,\,\,\,\,\,\,\,\,\,\,\,\,\,\,\,\,\,\,\,\,\,\,\,\,\,\,\,\,\,+ c\,c_g\,T^2s\left(\frac{\mu\,n}{\omega_m}(f_1\mu-e_1)-wT(\frac{1}{\omega_m}-\frac{1}{\omega_e})f_2-\frac{\mu\,n T f_2}{\omega_m}\right) \biggr\}.\nonumber
\end{align}
Let us now compute the heat current induced in the system due to turning the external sources on. Deriving the structure of the heat current is a little tricky. In \cite{Hartnoll:2009sz} and \cite{Herzog:2009xv} such current was identified in $2+1$ dimensional system as 
\begin{equation}\label{heat}
Q^i=T^{0i}-\mu j^i.
\end{equation}
In \cite{Abbasi:2018zoc} we derived the same formula for a $3+1$ dimensional system in the presence of the anomalies. The important point with the derivation of the  \cite{Abbasi:2018zoc} is that $T^{0i}$ and $J^i$ in  \eqref{heat} turns out to be the components of \textbf{covariant} stress tensor and \textbf{covariant} current\cite{Jensen:2012kj}. Recalling that the constitutive relations of the hydrodynamics are covariant under gauge and diffeomorphism transformations, we use \eqref{deltaTtz} and \eqref{delta_j_z} to write the linearized heat current as 
\begin{align}
\delta j^{z}_{th}=s\delta u_z+c_gB\,T
\delta T+\sigma_E\,\mu\,(\delta E_z+\mu\,\zeta_z)
\end{align}
Substituting \eqref{delta_T}, \eqref{delta_mu} and \eqref{delta_u}, we arrive at the heat current with the from given in \eqref{generalized_Ohm}. We then read the thermoelectric and thermal conductivity as the following
\begin{align}\label{alpha2}
\alpha_2=&-\mu\,\sigma_E+\frac{i}{\omega_m}\frac{Tn\, s}{w}\\\nonumber&
+\frac{i B^2}{(e_2f_1-e_1f_2)w^2}\bigg\{c^2\,\mu \,s\left(w(\frac{\mu f_2}{\omega_e}-\frac{e_2}{\omega_c})-\frac{\mu \, n}{\omega_m}(\mu f_2-e_2)\right)-c_g^2\frac{\,T^3\mu\, n^2f_1}{\omega_m}\,\\\nonumber&
\,\,\,\,\,\,\,\,\,\,\,\,\,\,\,\,\,\,\,\,\,\,\,\,\,\,\,\,\,\,\,\,\,\,\,\,\,\,\,\,\,\,\,\,\,\,\,\,\,\,\, + c\,c_g\,T\mu\, n\bigg(-\frac{s T^2f_2}{\omega_m}+w(\frac{\mu f_1}{\omega_e}-\frac{e_1}{\omega_c})-\frac{\mu\, n}{\omega_m}(\mu f_1-e_1)\bigg) \bigg\}\nonumber
\end{align}
\begin{align}\label{kappa}
\kappa=&\mu^2\,\sigma_E+\frac{i}{\omega_m}\frac{s^2}{w}\\\nonumber
&+\frac{i B^2}{(e_2f_1-e_1f_2)w^2} \biggl\{c^2\,\frac{T^2\mu^2s^2}{\omega_m}(e_2-f_2\mu)+c_g^2\,\,T^3\mu\, n \left(w(\frac{1}{\omega_e}-\frac{1}{\omega_m})+\frac{\mu \,n}{\omega_m}\right)f_1\\\nonumber
&
\,\,\,\,\,\,\,\,\,\,\,\,\,\,\,\,\,\,\,\,\,\,\,\,\,\,\,\,\,\,\,\,\,\,\,\,\,\,\,\,\,\,\,\,\,\,\,\,\,\,\,\,\,\,\,\,\,-c\,c_g\,T^2\mu\,s\left(T\,w\,f_2(\frac{1}{\omega_m}-\frac{1}{\omega_e})-\frac{\mu}{\omega_m}(e_1+T\,f_2-\mu\,f_1)\right)\biggr\}.
\end{align}
The common point among the four conductivities obtained above is that 
all of them quadratically depend on $B$ in the hydrodynamic limit $B\ll T^2$. One also observes that the megneto-transport happens through both
the chiral and mixed gauge-gravitational anomaly effects.
Moreover, not only the momentum relaxation contribute to the transport\footnote{Interestingly, In the non-magnetic part ($B=0$), all of the transport coefficients just depend on the $\tau_m$ coefficient corresponding to the momentum relaxation in the system. This is in complete agreement with what was obtained in \cite{Hartnoll:2007ih}. }, like what happans in an ordinary metal, but the relaxation in charge and energy do so. 

An important feature of the result obtained above is that the Onsager reciprocal relation, i.e. $\alpha_1=\alpha_2$, apparently fails, although the time reversal symmetry of correlation functions is still hold (See Appandix \ref{App_Onsager}.). To investigate it more, in the next subsection we use the conductivity formulas obtained in this section to compute the transport coefficients in a relativistic Weyl fluid in the limit $\mu \gg T$. The results will then be comparable with those of the chiral kinetic theory. We show that the Onsager relation will be satisfied if and only if the relaxation process in all channels happens at the same rate.

\section{  Weak regime: hydrodynamic magneto-transport in chiral kinetic theory}\label{comparison_hydro_CKT}
In 	\cite{Abbasi:2018zoc}, we studied the magneto-transport in a system of free Weyl fermions in the presence of a weak  magnetic field, $B\ll T^2$. By some motivations from Weyl semimetals, we firstly computed the second order quantum corrections to the dispersion of Weyl particles in the magnetic field. Then by developing a scheme to compute the phase space integrals, we analytically computed the thermodynamic quantities in the system  in low temperature limit, i.e. $T\ll \mu$:
\begin{align}\label{T_00}
\epsilon&=T^4 \left(\frac{\mu^4}{8 \pi ^2 T^4}+\frac{\mu^2}{4 T^2}+\frac{7 \pi ^2}{120}\right)+
\frac{e^2\text{B}^2}{24 \pi^2}-\left(log \frac{\mu}{\Delta_{\text{B}}}-\frac{\pi^2}{6}\frac{T^2}{\mu^2}\right)\frac{e^2\text{B}^2}{16 \pi^2}\\\label{p}
p&=T^4 \left(\frac{\mu^4}{24 \pi ^2 T^4}+\frac{\mu^2}{12 T^2}+\frac{7 \pi ^2}{360}\right)
+\frac{e^2\text{B}^2}{48 \pi^2}+\left(log \frac{\mu}{\Delta_{\text{B}}}-\frac{\pi^2}{6}\frac{T^2}{\mu^2}\right)\frac{e^2\text{B}^2}{16 \pi^2}\\\label{n}
n&=\,T^3 \left(\frac{\mu^3}{6 \pi ^2 T^3}+\frac{\mu }{6 T}\right)+\frac{e^2\text{B}^2}{16 \pi^2 \mu}\left(1+\frac{\pi^2T^2}{3\mu^2}\right).
\end{align}
In the above expressions $\Delta_{B}$ is an IR cutoff below which the chiral kinetic theory fails.

  As the next step in \cite{Abbasi:2018zoc}, in the framework of chiral kinetic theory and under the relaxation time approximation we found the longitudinal magneto-conductivities in the system
  \begin{align}\label{sigma_L}
  \sigma_{L}&=\,\frac{e^2 \tau}{3}\left(\frac{\mu^2}{2 \pi^2}+\frac{T^2}{6}+\frac{3 e^2 \text{B}^2}{16 \pi^2\mu^2}\left(1+\frac{\pi^2 T^2}{\mu^2}\right)\right)\\\label{alpha_L}
  T\,\alpha_{L}&=\,\frac{e\tau }{9} \, \mu T^2\left(1-\frac{3e^2B^2}{8\mu^4}\right)\\\label{kappa_L}
  T\, \kappa_{L}&=\,\frac{\tau (\pi T)^2}{9}\left(\frac{\mu^2}{2\pi^2}+\frac{7T^2}{10}+\frac{3\, e^2 \text{B}^2}{16\pi^2\mu^2}\right).
  \end{align}
  Analogous to the WSM case, our results showed a quadratic positive contribution to electrical and thermal conductivities.
  
  As an interesting point in \cite{Abbasi:2018zoc}, by deriving the Ward identities between one- and two-point functions in the infinite wave-length regime in the system, we found two constraint equations relating the magneto-conductivities in the presence of the anomalies as the following
  \begin{align}\label{alpha_zz}
  T\,\alpha_{L}&=\,-\frac{e\,n}{i \omega}-\,\frac{\mu}{e}\frac{}{}\, \sigma_{L}\\\label{kappa_zz}
  T\,\kappa_{L}&=\,-\frac{\epsilon+p -2\mu n}{i \omega}+\,\frac{\mu^2}{e^2}\, \sigma_{L}.
  \end{align}
   We then showed that the second order quantum correction to the kinetic energy dispersion was necessary for the magneto-conductivities to obey the mentioned constraint equations.

    \subsection{Constraints from time-reversal symmetry}
    \label{constraints}
 In this section we would like to show that how the hydrodynamic magneto-conductivities found in previous section may reproduce the
kinetic theory results found in \cite{Abbasi:2018zoc} \footnote{In \cite{Abbasi:2018zoc} we also argued that the same results for conductivities could be found directly in the hydrodynamic regime.}.
Let us recall that the transport coefficients we found by the hydrodynamic computations are general in the sense that we did not assume any specific thermodynamic equation of state to obtain them.
Now we want to apply the hydrodynamic results to the system of free Weyl fermions with the thermodynamics given by \eqref{T_00}, \eqref{p} and \eqref{n}.
Finding the thermodynamic derivatives $f_1$, $f_2$, $e_1$ and $e_2$ introduced in \eqref{epsilon_thermo} and \eqref{n_thermo}, the magneto-conductivities (\eqref{sigma}, \eqref{alpha1}, \eqref{alpha2} and \eqref{kappa}) in the $\mu\gg T$ limit and  $\omega=0$ then read 
\begin{align}\label{sigma_hydro}
\sigma=&\sigma_E-\frac{\mu^2\,\tau_m}{6\pi^2}\big(1+\frac{24\pi^4}{45}(\frac{T}{\mu})^4-\frac{48\pi^6}{45}(\frac{T}{\mu})^6\big)\\\nonumber&
\,\,\,\,\,\,\,\,\,\,\,\,\,\,\,\,\,\,\,\,\,\,\,\,\,\,\,\,\,\,\,\,\,\,\,\,\,\,\,\,\,\,\,\,\,\,\,\,\,\,\,\,+\frac{B^2}{16\pi^2\mu^2}(2\tau_c-2\tau_e+\tau_m)+\frac{B^2 \,T^2}{\mu^4}(\frac{13}{40}\tau_c-\frac{7}{60}\tau_e-\frac{1}{8}\tau_m) \\\label{alpha_1_hydro}
\alpha_1=&-\mu\,\sigma_E+\frac{\mu\,T^2\tau_m}{6}\bigg(1-\frac{24\pi^2}{45}(\frac{T}{\mu})^2+\frac{48\pi^4}{45}(\frac{T}{\mu})^4\bigg)-\frac{B^2\,T_0^2}{48\mu^3}(2\tau_e+\tau_m)\\\label{alpha_2_hydro}
\alpha_2=&-\mu\,\sigma_E+\frac{\mu\,T^2\tau_m}{6}\bigg(1-\frac{24\pi^2}{45}(\frac{T}{\mu})^2+\frac{48\pi^4}{45}(\frac{T}{\mu})^4\bigg)\\\nonumber&
\,\,\,\,\,\,\,\,\,\,\,\,\,\,\,\,\,\,\,\,\,\,\,\,\,\,\,\,\,\,\,\,\,\,\,\,\,\,\,\,\,\,\,\,\,\,\,\,\,\,\,\,+\,\frac{B^2}{8\pi^2\mu}(\tau_e-\tau_c)+\frac{B^2T^2}{\mu^3}(\frac{7}{48}\tau_m+\frac{7}{60}\tau_e-\frac{13}{40}\tau_c)\\\label{kappa_hydro}
\kappa=&\mu^2\,\sigma_E+\frac{\pi^2T^4\,\tau_m}{6}\big(1-\frac{48\pi^2}{45}(\frac{T}{\mu})^2\big)+\frac{ B^2 \,T^2 \tau_e}{24\mu^2}.
\end{align}
As mentioned before, the non-magnetic part of the magneto-conductivities just depends on the  momentum relaxation, i.e. $\tau_m$. While in agreement with the Onsager relations the non-magnetic part of $\alpha_1$ and $\alpha_2$ are the same \cite{Hartnoll:2007ih,Lucas:2017idv},  their magnetic part seems to differ from each other in general. 
Demanding the Onsager reciprocal relation is valid when the system is described with the quasi-hydrodynamics, we obtain the following two equations:
\begin{equation}
\begin{split}
O(\frac{1}{\mu}):&\,\,\,\,\,\,\,\,\,\,\,\,\,\,\,\,\,\,\,\,\,\,\,\,\,\,\,\,\,\,\,\,\,\,\,\,\,\,\,\,\,\,\,\,\,\,\,\,\,\,\,\,\,\,\,\tau_e-\tau_c=0\\
O(\frac{1}{\mu^3}):&\,\,\,\,\,\,\,\,\,\,\,\,\,\,\,\,\,\,\,\,\,\frac{7}{48}\tau_m+\frac{7}{60}\tau_e-\frac{13}{40}\tau_c=-\frac{1}{48}(2 \tau_e+\tau_m)
\end{split}
\end{equation} 
We so arrive at
\begin{equation}
\boxed{\tau_e=\tau_m=\tau_c}.
\end{equation}
This result simply says that in order for the conductivities to be Onsager reciprocal, charge, momentum and energy cannot relax at different rates. The common rate at which they dissipate is characterized with just one relaxation time parameter, e.g. $\tau$, which is the analogous to the parameter we used in the kinetic theory computations in \cite{Abbasi:2018zoc}. \textbf{If the latter does not happen, the time reversal symmetry of the correlation functions leads the constraints on the conductivities to be modified.} See Appendix \ref{App_Onsager} for more details.

At this point, one may be tempted to compare the conductivities given in equations \eqref{sigma_hydro} to \eqref{kappa_hydro} with their counterpart in the kinetic theory side. Let us denote that although we were able to compute the thermodynamic derivatives in the Weyl fluid, we did not however find the corresponding quantum conductivity $\sigma_E$. So, to compare the hydrodynamic results with the kinetic theory ones, we do as the following. We equate each hydrodynamic conductivity with its counterpart to find $\sigma_E$ and consequently find four equations. If the hydrodynamics precisely coincides with the kinetic theory at the limit $B\ll T^2\ll \mu^2$, one expects all the four equations give the same result for $\sigma_E$.

Interestingly, by considering $\tau_e=\tau_m=\tau_c\equiv\tau$, all the equations
\begin{equation*}
\begin{split}
\eqref{sigma_hydro}&=\eqref{sigma_L},\,\,\,\,\,\,\,\,\,\,\,\,\,\,\,\,\,\,\,\,\eqref{alpha_1_hydro}=\eqref{alpha_L},\\
\eqref{alpha_2_hydro}&=\eqref{alpha_L},\,\,\,\,\,\,\,\,\,\,\,\,\,\,\,\,\,\,\,\,\eqref{kappa_hydro}=\eqref{kappa_L}
\end{split}
\end{equation*}
give the same result:
\begin{align}\label{sigma_E_kinetic}
\boxed{
\sigma_E=\frac{T^2\,\tau}{18}\bigg(1-\frac{8\pi^2}{5}(\frac{T}{\mu})^2+\frac{16\pi^4}{5}(\frac{T}{\mu})^4\bigg)-\frac{B^2 \tau}{48\mu^2}\frac{T^2}{\mu^2}}.
\end{align}
Let us recall that as mentioned in \cite{Hartnoll:2016apf} the computation of $\sigma_{E}$ in general cannot be reliably computed in the quasi-hydrodynamics developed in the current paper. However we see that combing the quasi-hydrodynamic with the kinetic theory results of a free Weyl fluid gives the value of $\sigma_{E}$ as  \eqref{sigma_E_kinetic}.
This expression not only may be regraded as a new result related to relativistic Weyl fluid, but it confirms that the uniform and time independent state of the system studied in the kinetic theory in \cite{Abbasi:2018zoc} had been indeed in the hydrodynamic limit.


\subsection{Getting back to the hydrodynamics}
\label{gettin_back}
In the previous subsection we showed that in an anomalous system with weakly broken symmetries, the Onsager reciprocal relations can still hold if the relaxation in all channels occurs at the same rate.
From now on we focus on this case and reconsider the hydrodynamic results of \sec{magneto_transport_from_hydro} under the assumption $\tau_e=\tau_m=\tau_c\equiv\tau$. Let us first rewrite the longitudinal electrical conductivity \eqref{sigma} in this limit
\begin{equation}
\sigma=\sigma_{E}+\frac{i}{\tilde{\omega}}\frac{n^2}{w}+\frac{i}{\tilde{\omega}}\frac{B^2}{ w^2 (e_2 f_1-e_1 f_2)}\bigg\{c^2 (e_2-f_2 \mu)T^2 s^2+ c c_g T(e_1+f_2 T-f_1 \mu)n T s+c_g^2 f_1 T^3 n^3\bigg\}
\end{equation}
with $\tilde{\omega}=\omega+i/\tau$. Interestingly when rewriting \eqref{alpha1}, \eqref{alpha2} and \eqref{kappa} in the same limit, we find ($\alpha_1=\alpha_2=\alpha$)
\begin{align}\label{ward_1}
T\,\alpha&=\,-\frac{i}{\tilde{ \omega}}\,n-\,\mu\, \sigma\\\label{ward_2}
T\,\kappa&=\,-\frac{i}{ \tilde{ \omega}}(\epsilon+p -2\mu n)+\,\mu^2\, \sigma.
\end{align}
These are nothing other than the constraint equations were found in \cite{Abbasi:2018zoc} from the Ward identity arguments (See the details around \eqref{alpha_zz}.). Compared to the latter, here, $\omega$ has been replaced with $\tilde{ \omega}=\omega+i/\tau$. Since the above constraints were originally found from the (non-)covariance under gauge and diffeomorphism transformations, the appearance of  $\tilde{ \omega}$ suggests that there might be modified charge and diffeomorphism transformations by which, one can derive the equations \eqref{E1}, \eqref{E2} and \eqref{E3} from a generating functional. Finding such transformations is out of the scope of this paper \cite{Abbasi:2019zoc}.

Before ending this subsection, let us recall that as it was mentioned in \cite{Landsteiner:2014vua}, even in the zero density limit $n=0$, the magneto electrical conductivity is still non-vanishing. In this limit $f_2=0$ and $w=Ts$, so one arrives at
\begin{equation}
\sigma=\sigma_E+\,\frac{\tau}{1-i \omega \tau}\,\frac{B^2 c^2}{\big(\partial n/\partial \mu\big)_T}
\end{equation}
In the next section we evaluate the above conductivity in a holographic system and compare the results with that of found in \cite{Landsteiner:2014vua}.

\section{Strong regime: holographic magneto-transport}
\label{2}
In \sec{hydro} we studied the magneto-transport in a general quantum system in the hydrodynamic limit. Then in \sec{comparison_hydro_CKT} by putting the thermodynamic equation of state of the single chirality relativistic Weyl fluid (found from chiral kinetic theory) into the hydro results, we computed the transport coefficients in this specific system in $\mu\gg T$.   How about the opposite limit, namely $\mu \ll T$?   Such limit might be interesting in the study of quark-gluon-plasma produced in the heavy ion scattering experiments. However, there are evidences showing that the mentioned system is a strongly coupled system 	\cite{Shuryak:2003xe} in this limit. Alternatively, one may study the problem in the context of holography  by considering the dual gravity of strongly coupled gauge theory \cite{Maldacena:1997re,Witten:1998qj}.  
 In \cite{DHoker:2009ixq} in the context of holography, the gravity dual for a 
strongly coupled boundary chiral gauge theory in the presence of a constant magnetic field has been introduced. In this section we follow the method of \cite{DHoker:2009ixq} and explicitly derive the bulk solution as well as the thermodynamic quantities of the boundary  theory
in the limit $\mu\ll T$. Then we would be able to compute the conductivities in  the hydrodynamic limit.

\subsection{Gravity set-up}
The gravity dual of 3+1 dimensional chiral gauge theory at finite chiral density and in the presence of  a background magnetic field is a 4+1 dimensional Einstein/Maxwell theory with a Chern-Simones term\footnote{In this section we do not consider the effects of mixed gauge-gravitational anomaly in the gauge theory.}. While the value of the Chern-Simones coupling $k$ captures the strength of the anomaly in the boundary theory, we keep it general without considering any specific value for it \cite{DHoker:2009ixq}. The action for such theory is written as 
\begin{equation}
S = - \frac{1}{16 \pi G_5} \int_{\mathcal{M}} d^5 x \,\,\sqrt{-g} \left(R - \frac{12}{L^2} + F^{M N} F_{M N}\right)+S_{CS}+ S_{bdy}
\end{equation}
with the Chern-Simons action being as the following
\begin{equation}
S_{CS}=\frac{k}{12 \pi G_5}\int A \wedge F \wedge F=\frac{k}{192 \pi G_5}\int d^5x \sqrt{-g}\epsilon^{MNPQE}A_{M}F_{NP}F_{QR}
\end{equation}
The appropriate boundary term which is needed for the purpose of the renormalizing boundary quantities is given by\footnote{This form of the boundary term is specific to the case in which $g_{r\mu}$ vanishes asymptotically}
\begin{equation}
S_{bdy}= -\frac{1}{8 \pi G_5} \int_{\partial \mathcal{M}} d^4 x \,\,\sqrt{-\gamma} \left(\mathcal{K} - \frac{3}{L} + \frac{L}{4} R(\gamma) + \frac{L}{2} \ln(\frac{r}{L}) F^{\mu \nu} F_{\mu \nu}\right)
\end{equation}
with $\gamma_{\mu \nu}$ being the induced  metric on the boundary and $\mathcal{K}$ is the trace of the boundary extrinsic curvature. Let us denote that in what follows, we refer to the bulk coordinates with Latin indices $x^{M}=\{x^{\mu},r\}$ while to the boundary coordinates with the Greek ones $\mu=0,1,2,3$. We also set the radius of AdS to unity $L=1$. 

In addition to the Bianchi identity $dF=0$, the field equations are as the following
\begin{align}\label{Maxwell}
0&= d * F+k\, F \wedge F\\\label{Einstein}
R_{MN}&=4 g_{MN}+\frac{1}{3}g_{MN} F^{AB}F_{AB}-2 F_{MP}F_{N}^{\,\,P}
\end{align}
In order to make an appropriate ansatz for the bulk fields let us recall that the presence of a uniform background magnetic filed, say directed in $x^3$ direction, simply reduces the full rotational symmetry on the boundary theory to a rotational symmetry in the spacial plane perpendicular to the magnetic field. Since the translation symmetry of $x^{\mu}$ is still hold, the ansatz may be given by   
\begin{equation}\label{metric}
ds^2=\frac{dr^2}{U(r)}-U(r)dt^2+ e^{2V(r)}(dx_1^2+dx_2^2)+e^{2W(r)}(dx_3+C(r)dt)^2
\end{equation}
for the metric field and by
\begin{equation}\label{field_strenght}
F=E(r) dr\wedge dt+B dx_1\wedge dx_2+ P(r) dx_3\wedge dr 
\end{equation}
for the Maxwell field strength.
In the following subsections we will solve the field equations to find the unspecified functions in  \eqref{metric} and \eqref{field_strenght}, perturbatively by a double expansion.  At small $B$, the perturbative solution were found in \cite{DHoker:2009ixq} before,  by taking the ansatz
\begin{align}
U&= U_0 + B^2 U_2 \,\,\,\,\,\,\,\,\,\,\,\,\,\,\,\,\,\,\,\,\,\,E = E_0 + B^2 E_2\nonumber\\\label{asymptotic}
W &= W_0 + B^2 W_2\,\,\,\,\,\,\,\,\,\,\, \,\,\,\,\,\,\,\,\, C = C_0 + B C_1\\
 V &= V_0 + B^2 V_2 \,\,\,\,\,\,\,\,\,\,\,\,\,\,\,\,\,\,\,\, \,\,\,\,\,P = P_0 + B P_1\nonumber
\end{align}
In the above expressions,  the functions with subscript $0$ correspond to a AdS5 Reissner-Nordstorm black hole in the absence of the magnetic field.
The correction functions with subscripts $1$ and $2$ (depending on whether the quantity is odd or even under charge conjugation) have been given in terms of some unevaluated integrals in \cite{DHoker:2009ixq}. We implement the second perturbation and explicitly evaluate the correction functions to leading order in  the "chemical potential to temperature ratio" of the boundary theory, namely $\mu/(\pi T)$.

\subsection{Solution at $B=0$ and small $\mu_0/(\pi T_0)$}
In the absence of the magnetic field, the solution \eqref{metric} and \eqref{field_strenght} simply reduces to a five dimensional AdS-RN black hole with
$P_0= C_0=0$,
\begin{equation}\label{RN}
E_0= \frac{Q}{r^3},\,\,\,\,\,
V_0=W_0=\ln r,\,\,\,\,\, U_0= r^2 + \frac{Q^2}{3 r^4} - \frac{M}{r^2}.
\end{equation}
The radii of the inner and outer horizons, namely $r_{-}$ and $r_{+}$ are the roots of equation $U_{0}=0$. One finds
\begin{align}
\frac{Q^2}{3}=r_+^2 r_-^2 (r_+^2+ r_-^2),\,\,\,\,M= r_+^4 +r_-^4 + r_+^2 r_-^2. 
\end{align}
It is worth-mentioning that the whole solution is fixed once two parameters $M$ and $Q$ are given. Alternatively, one may describe the boundary physics by another pair of parameters, e.g. $r_{+}$ and $r_{-}$. A more physical choice which might be useful to study the boundary field theory is the pair of temperature and  chemical potential. Using the solution \eqref{RN}, the black hole temperature and chemical potential are then given in terms of the $r_{\pm}$
\begin{align}
T_0\equiv\,\frac{U'_{0}(r_{+})}{4 \pi} &= \frac{r_+}{2 \pi} \left(2 - \left(\frac{r_-}{r_+}\right)^2 - \left(\frac{r_-}{r_+}\right)^4\right),\\
\mu_0 \equiv\,\, A_{t}\big|_{r_{+}}&= \frac{Q}{2r_+^2}.
\end{align} 
 In order to rewrite all the metric and field strength components in terms of $T_{0}$ and $\mu_{0}$, it is needed to inverse the above two equations and find $r_{+}$ and $r_{-}$ in terms of $T_0$ and $\mu_0$. To do this analytically, at this point we enter the second perturbation, namely the expansion over $\mu_0/(\pi T_0)$. We find 
\begin{align}
r_+ =\, \frac{\pi T_0}{2} \left(1+ \sqrt{1+ \frac{2 \mu_0^2}{3 \pi^2 T_0^2}}\right)& =\, \pi T_0 \left(1+ \frac{\nu_0^2}{6}\right) + \mathcal{O}(\nu_0^4),\\
r_{-} =\, \frac{r_{+}}{\sqrt{2}} \sqrt{-1 + \sqrt{9-\frac{8\pi T_0}{r_{+}}}}& =\,\pi T_0 \frac{\nu_0 }{\sqrt{3}} + \mathcal{O}(\nu_0^3)
\end{align}
where we have defined $\nu_0=\mu_0/(\pi T_0)$.
In the next subsection we show how the above perturbation helps us to analytically find the first non-trivial correction of the magnetic field to the metric and field strength.
\subsection{Perturbative solution at small $B$ and small $\mu_0/\pi T_0$}
Solving the equations \eqref{Maxwell} and \eqref{Einstein} perturbatively in small $B$, the unspecified functions of the metric introduced in \eqref{asymptotic} may be given as the following
\begin{align}
S_2(r)& = 2 \int_{\infty}^{r} dr' \left(\frac{1}{r} - \frac{1}{r'}\right) P_1(r')^2,\nonumber\\
 E_2(r)& = -\frac{Q}{r^3} S_2(r) - P_1(r) C_1(r) - \frac{2 k}{r^3} \int_{\infty}^{r} dr' P_1(r'),\nonumber\\
U_2(r) &= \int_{\infty}^{r} \frac{dr''}{r''^3} \int_{r_+}^{r''} dr' X(r') - \frac{a_3}{2 r^2},\nonumber\\
T_2(r) &= \int_{\infty}^{r} dr'' \frac{1}{r''^3 U_0(r'')} \int_{r_+}^{r''} dr' \left(\frac{1}{2} r'^5 (\frac{d C_1(r')}{dr'})^2 +2 r' U_0(r') (P_1(r'))^2- \frac{2}{r'}\right),
\end{align}
with 
\begin{equation}
X(r) = - r^3 \frac{d U_0(r)}{dr} \frac{d S_2(r)}{dr} + \frac{16 Q}{3} \left(E_2(r) + C_1(r) P_1(r)\right) + r^5 (\frac{d C_1(r)}{dr})^2 + \frac{4 r U_0(r) P_1(r)^2}{3} + \frac{4}{3 r},\nonumber
\end{equation}
and
\begin{equation}
V_2(r) = \frac{S_2(r) +  T_2(r)}{3}, \,\,\,\,\,\,\, W_2(r) = \frac{S_2(r) - 2 T_2(r)}{3}.
\end{equation}
The unspecified component of the filed strength are given by
\begin{align}
C_1(r) &= - k Q^2 \frac{U_0(r)}{r^2} \int_{\infty}^{r} \frac{dr'}{r' U_0^2(r')} (\frac{1}{r'^2} - \frac{1}{r_{+}^2})^2,\nonumber\\
P_1(r)& = \frac{\rho}{r U_0(r)} \left(C_1(r) + \frac{k}{r^2} - \frac{k}{r_+^2}\right),\nonumber
\end{align}
When expanding the integrands in terms of $\nu_0=\mu_0/(\pi T_0)$, the above expressions can be evaluated analytically. We find
\begin{align}
C_1(r)&=-\frac{2 k \nu_0^2}{r^2}\left(1-\frac{(\pi T_0)^2}{r^2}\right)+\frac{2 k \nu_0^2}{(\pi T_0)^2}\left(1-\frac{(\pi T_0)^4}{r^4}\right)\ln\left(1+\frac{(\pi T_0)^2}{r^2}\right)\\\nonumber
P_1(r)&=-\frac{2 k\, (\pi T_0)\,\nu_0}{r(r^2+(\pi T_0)^2)}\\\nonumber
S_2(r)&=\frac{2 k^2\, \nu_0^2}{(\pi T_0)^4\, r^2}\left[(\pi T_0)\big(3 r \pi-2 (\pi T_0)\big)- 6r (\pi T_0) \cot^{-1}\frac{\pi T_0}{r}-2r \ln\left(1+\frac{(\pi T_0)^2}{r^2}\right)\right]\nonumber
\end{align}
together with
\begin{align}\nonumber
E_2(r)&=-\frac{2 k^2\, \nu_0}{\pi T_0\,r^3 }\ln\left(1+\frac{(\pi T_0)^2}{r^2}\right)\\ \label{metric_corrections}
 U_2(r) &=  - \frac{1}{3 r^2}\left(2 \ln \frac{r}{\pi T_0} +1\right) - \frac{a_3}{2 r^2}+ \nu_0^2\, \mathcal{F}(r).
\end{align}
Where we have promoted the $a_3$ constant in $U_2$ to be a function of boundary thermodynamic data. 
Note that since we do not need $T_2(r)$ in our later computations, we have not written its expanded expression in the above. In the above expressions $\mathcal{F}(r)$
is given by
\begin{align}
\mathcal{F}(r) = & \frac{4 k^2 (\pi T_0)^2}{3 r^4} \left[1 - 2 \ln \left(1 + \frac{\pi^2 T_0^2}{r^2}\right)\right]+ \frac{4 k^2 (\pi T_0)}{r^3}\left[ \pi - 2 \tan^{-1}\left(\frac{r}{\pi T_0}\right)\right] \\\nonumber
& + \frac{ 2}{ r^2} \left[\frac{2}{9} + k^2 \left(4 \ln 4- \pi - \frac{16}{3} \ln \left(1 + \frac{\pi^2 T_0^2}{r^2}\right) \right)\right] + \frac{4 k^2}{\pi^2 T_0^3} \left[r \left(1- \frac{2}{\pi} \tan^{-1}\big(\frac{r}{\pi T_0}\big)\right) - 2 T_0\right].
\end{align}
\subsection{Stress tensor and charge current on the boundary}
In order to  compute the physical quantities in the boundary field theory from the  bulk informations, we have to find the asymptotic behavior of the solutions found in the previous subsection. The solutions at $r \rightarrow \infty$ read
\begin{align}\nonumber
C_1&= \frac{k (\pi T_0)\,\nu_0^2}{r^4},\,\,\,\,\,\,\,\,\,\,\,\,\,\,\,\,\,\,\,\,\,\,\,\,\,\,\,P_1= -\frac{2k (\pi T_0)\,\nu_0}{r^3},\,\,\,\,\,\,\,\,E_2 = - \frac{2 k^2 (\pi T_0) \,\nu_0}{r^5}\\\nonumber
U_2 &= - \frac{1}{3r^2}\left(1+2 \ln \frac{r}{\pi T_0}-\frac{4k^2}{3}+k^2\nu_0^2(8+6\pi+48\ln 2-18 \ln (\pi T_0))+\frac{3 a(T_0,\nu_0)}{2}\right),\\
S_{2}&=-\frac{4 k^2 (\pi T_0)^2\,\nu_0^2}{15 r^6},\,\,\,\,\,\,\,\,\,\,\,\,\,\,\,\,\,\,T_2=\frac{1}{8r^4}\left(1+\ln\frac{r^4}{(\pi T_0)^4}\right)-\frac{\nu_0^2}{3r^4}(1-3k^2+6k^2 \ln 2)
\end{align}
At this point everything seems ready to compute the boundary stress tensor and charge current. To proceed, we follow \cite{Balasubramanian:1999re} and write
\begin{equation}\label{AdS_dictionary}
8 \pi G_5 T^{\mu \nu}= r^2 \left\{\mathcal{K}^{\mu \nu}-  \mathcal{K} \gamma^{\mu \nu}   - 3 \gamma^{\mu \nu} -\frac{1}{2} G^{\mu \nu}(\gamma) - 2 \ln \frac{r}{r_0} (F^{\mu \alpha} F^{\nu}_{\alpha} -\frac{1}{4} \gamma^{\mu \nu} F^{\alpha \beta} F_{\alpha \beta})\right\}
\end{equation}
where $\mathcal{K}_{\mu\nu}$ is the boundary extrinsic curvature  (see Appendix \ref{extrinsic} for more detailed computations)
Here in contrast to \cite{DHoker:2009ixq}, we have taken $\ln\frac{r}{r_0}=\ln\left(\frac{r}{L}\frac{L}{r_0}\right)$ instead of $\ln\frac{r}{L}$ in the counter term within the expression of energy-momentum tensor.  This modification not only keeps the energy-momentum on the boundary finite, but makes a finite shift $\sim \ln r_0$ in its diagonal components which turns out to be necessary to find physical results on the boundary. Equivalently we would rather to work with an energy scale $\Delta$ in our computations, as $r_0=L^2 \Delta$. So when we compute $\ln \frac{r}{r_0}$ at  a typical radius in the bulk like $r_{+}=L^2 \pi T_0$, we obtain a logarithm with the form $\ln\frac{\Delta}{\pi T_0}$ which will be meaningful if $\Delta$ is a definite energy scale of the boundary theory.  Let us denote that we temporarily restored the AdS radius $L$ in this discussion. In the following we take $L=1$ as before.  

From \eqref{AdS_dictionary} we then find the non-vanishing components of the stress tensor 
\begin{align}\nonumber
T_{00}=&\frac{1}{16\pi G_5}\big(3(\pi T_0)^4+12 (\pi T_0)^2\mu_0^2+8 \mu_0^4\big)\\\nonumber
&\,\,\,\,\,\,+\frac{B^2}{16 \pi G_5}\bigg(1+\frac{3a_3(T_0,\mu_0)}{2}+2 \ln \frac{\Delta}{\pi T_0}\bigg)+\frac{B^2 \nu_0^2}{2\pi G_5}\left(k^2(\frac{3\pi}{4}+1-6 \ln 2)-\frac{1}{6}\right)\\\nonumber
T_{33}=&\frac{1}{16\pi G_5}\big((\pi T_0)^4+4 (\pi T_0)^2\mu_0^2+\frac{8}{3} \mu_0^4\big)\\\nonumber
&\,\,\,\,\,\,-\frac{B^2}{16 \pi G_5}\bigg(\frac{1}{3}+\frac{3a_3(T_0,\mu_0)}{2}+2 \ln \frac{\Delta}{\pi T_0}\bigg)+\frac{B^2 \nu_0^2}{2\pi G_5}\left(k^2(\frac{\pi}{4}-\frac{1}{3}-\frac{2}{3} \ln 2)+\frac{1}{6}\right)\\\nonumber
T_{ii}=&\frac{1}{16\pi G_5}\big((\pi T_0)^4+4 (\pi T_0)^2\mu_0^2+\frac{8}{3} \mu_0^4\big)\\\label{T_before_cor}
&\,\,\,\,\,\,-\frac{B^2}{16 \pi G_5}\bigg(\frac{1}{3}+\frac{3a_3(T_0,\mu_0)}{2}-2 \ln \frac{\Delta}{\pi T_0}\bigg)+\frac{B^2 \nu_0^2}{2\pi G_5}\left(k^2(\frac{\pi}{4}+\frac{2}{3}-\frac{8}{3} \ln 2)-\frac{1}{6}\right)
\end{align}
with
\begin{equation}\label{T_03}
T_{03}=\frac{1}{4\pi G_5}\,B \,k \,\mu_0^2.
\end{equation}
A very important point with the above results is that $T_0$ and $\mu_0$ appearing in the expressions are not the physical temperature and chemical potential of the boundary theory! The reason is that these quantities were computed form the metric and field strength solutions of the bulk before the magnetic field came to perturb them. From \eqref{asymptotic} it is obvious that the location of the horizon in the bulk is affected due to the magnetic field. When $B=0$, the horizon of RN black hole is the root of $U_0(r)=0$ while when the brane is magnetized, the radius of horizon will be the perturbative root of $U_0(r)+B^2 U_2(r)=0$. The chemical potential relatively will be changed. 

Before computing the correction to the temperature and chemical potential, let us first find the components of the boundary charge current   in terms of $T_0$ and  $\mu_0$ via the standard formula \cite{Balasubramanian:1999re}
\begin{equation}\label{Kraus_charge}
-4 \pi G_5 J^\mu = r^3 \gamma^{\mu \nu (0)} F_{r \nu} + \frac{k}{3} \epsilon^{\alpha \beta \gamma \mu} A_\alpha F_{\beta \gamma}.
\end{equation}
To evaluate it we need the explicit form of the  gauge field solution $A_{M}$ in the bulk. We integrate \eqref{field_strenght} and obtain  
\begin{align}\nonumber
A_0(r)& = \int_{\infty}^{r} dr' E(r') = - \frac{Q}{2 r^2} + \frac{k^2 Q B^2}{4 r_+^2 r^4} \\\nonumber
 A_3(r)& = - \int_{\infty}^{r} dr' P(r') = - \frac{k Q B}{2 r_+ r^2} ,\\\label{A_r}
 A_r(r)&=0,\,\,\,A_x = -\frac{1}{2} B y, ,\,\,\,A_y = \frac{1}{2} B x
\end{align}
Now from the formula \eqref{Kraus_charge} we find the non-vanishing components of the charge current
\begin{align}\label{J_before_cor}
J^0 = \frac{1}{2\pi G_5}(\pi T_0)^3\,(\nu_0+\frac{4}{3}\nu_0^3),\,\,\,\,\, J^3 =  -\frac{1}{2\pi G_5} B \,k\, \mu_0
\end{align}
In the next subsection, by finding the corrected location of the horizon we compute the corrected temperature and chemical potential. Having them, we will be able to fix the constant $a_3(T_0, \mu_0)$ and find the physical stress tensor and charge current on the boundary.

\subsection{Corrected horizon and physical boundary quantities}
As mentioned earlier, the corrections coming from the magnetic field displaces the horizon from being the root of $U_0=0$ to that of $U_0+B^2 U_2=0$. To find the corrected location of the horizons $\overline{r}_{\pm}$, let us take
\begin{align}
\overline{r}_{\pm}&=r_{\pm}+B^2\,\tilde{r}_{\pm}\\
T&=T_0+ B^2\, T_2\\
\mu&=\mu_0+ B^2\,\mu_2.
\end{align}
The correction term $\tilde{r}_{\pm}$ can be found by perturbatively finding the roots of the corrected blackening factor $U(r)=U_0(r)+B^2U_2(r)$ as the following 
\begin{equation}
U(\overline{r}_+)=0\,\,\,\,\rightarrow\,\,\,\,\,\,\tilde{r}_+ = - \frac{U_2(r_+)}{U_0'(r_+)}.
\end{equation}
Using the expression of $U_2(r)$ from the \eqref{metric_corrections}, we find
\begin{equation}\label{corrected_r_H}
\tilde{r}_{+}=\frac{1}{12 \pi^3T_0^3}\bigg(1+\frac{3}{2}a_{3}(T_0,\mu_0)\bigg)-\frac{\nu_0^2}{9\pi^3T_0^3}\bigg(1+\frac{3}{2}a_{3}(T_0,\mu_0)-15k^2(1-\frac{3}{10}\pi-\frac{4}{10}\ln 2)\bigg)
\end{equation}
Having found the location of the horizon, now we are able to find the corresponding Hawking temperature of the black hole 
\begin{align}\label{trans_T}
T=&\,  \frac{U_0'(\overline{r}_+)}{4 \pi} + B^2\frac{U_2'(r_+)}{4 \pi} 
= \frac{U_0'(r_+)}{4 \pi} \left(1 + B^2 \left(\frac{U_2(r)}{U_0'(r)}\right)'_{r=r_+}\right) \\\nonumber
=&T_0-\frac{B^2}{12 \pi^4 T_0^3}\big(1-\frac{3a_3(T_0,\mu_0)}{2}\big)+\frac{B^2 \,\nu_0^2}{2\pi^4 T_0^3}\left(1+k^2(\pi-\frac{10}{3}+\frac{4}{3}\ln 2)+\frac{a_3(T_0,\mu_0)}{2}\right)
\end{align}
Similarly, turning on the magnetic field changes the electric field in the bulk and consequently the chemical potential gets correction as it follows
\begin{align}\nonumber
\mu=\int_{\infty}^{\overline{r}_{+}}E(r)dr=&\int_{\infty}^{\overline{r}_{+}}\,(E_0(r)+B^2\,E_2(r))dr\\\label{trans_J}
=&\mu_0-\frac{B^2 \mu_0}{6 \pi^4 T_0^4}\left(1+\frac{3a_3(T_0,\mu_0)}{2}-6k^2(1-2\ln 2)\right).
\end{align} 
Let us denote that at the order of perturbation  we are studying the boundary theory, its corresponding temperature and chemical potential are $T$ and $\mu$, respectively. 
So once $T_0$ and $\mu_0$ are found in terms of $T$ and $\mu$ from \eqref{trans_T} and \eqref{trans_J}, we can use the ansatz
\begin{equation}\label{ansatz}
a_3(T, \mu)=a_0+\frac{\mu}{T}a_1(T)+\frac{\mu^2}{T^2}a_{2}(T).
\end{equation}
and rewrite the boundary stress tensor \eqref{T_before_cor} and charge current \eqref{J_before_cor} in terms of the physical temperature and chemical potential:  
\begin{align}\label{T00_a_3}
T_{00}=&\frac{1}{16\pi G_5}\big(3(\pi T)^4+12 (\pi T)^2\mu^2+8 \mu^4\big)\\\nonumber
&\,\,\,\,\,\,+\frac{B^2}{8 \pi G_5}\bigg(1- \ln \frac{\Delta}{\pi T}\bigg)+\frac{B^2}{2\pi G_5}\frac{\mu^2}{(\pi T)^2}\left(\frac{7}{2}\, k^2(1-2 \ln 2)-\frac{5}{12}-\frac{9}{8}a_3\right)\\\label{T33_a_3}
T_{33}=&\frac{1}{16\pi G_5}\big((\pi T)^4+4 (\pi T)^2\mu^2+\frac{8}{3} \mu^4\big)\\\nonumber
&\,\,\,\,\,\,,+\frac{B^2}{8 \pi G_5} \ln \frac{\Delta}{\pi T}\,+\frac{B^2}{2\pi G_5}\frac{\mu^2}{(\pi T)^2}\left(\frac{1}{2}\, k^2(1-2 \ln 2)+\frac{1}{12}-\frac{3}{8}a_3\right)\\\label{J0_a_3}
J^0 =& \frac{1}{2\pi G_5}(\pi T)^2\,\mu+\frac{B^2}{8\pi G_5}\frac{\mu}{(\pi T)^2}\left(1-\frac{3}{2}a_0-\frac{3}{2}\frac{\mu}{T}a_1\right)
\end{align}
Let us denote that due to our further requirements, we have only written $T^{00}=\epsilon$, $T^{33}=p$ and $J^0=n$. $T^{03}$
and $J^{3}$ do not get correction to our desired order and are still given by \eqref{T_before_cor} and \eqref{J_before_cor}, however, by replacing $T_0$ and $\mu_0$ with $T$ and $\mu$, respectively.

To fix the coefficient (functions) in \eqref{ansatz}, we now argue as it follows. The pressure $p=T^{33}$ and the charge density $n=J^0$ are constrained to satisfy the thermodynamic relation $n=(\partial p/\partial \mu)_{T}$. Using  \eqref{T33_a_3} and \eqref{J0_a_3}, this relation gives
\begin{equation}\label{a_0_a_1}
\boxed{
a_0=-\frac{2}{9}+\frac{8}{3} k^2(1-2 \ln 2),\,\,\,\,\,\,\,\,\,\,\,a_1=0,\,\,\,\,\,\,\,\,\,\,\,a_2=\text{undetermined}}
\end{equation}
There are two other thermodynamic relations which have to be satisfied. The \textbf{first} one is the Gibbs-Duhem relation. Following the magneto-thermodynamic discussion in  \cite{Grozdanov:2016tdf,Hernandez:2017mch}, the enthalpy density in our system, namely $w=\epsilon+p$, has to be given as
\begin{equation}\label{gibbs_duhem}
\epsilon+p=\mu\left(\frac{\partial p}{\partial \mu}\right)_{T}+T\left(\frac{\partial p}{\partial T}\right)_{\mu}.
\end{equation}
Using \eqref{T00_a_3}, \eqref{T33_a_3} and \eqref{J0_a_3}, one can show that \eqref{gibbs_duhem} is satisfied if and only if $a_0$ and $a_1$ are given by \eqref{a_0_a_1}. The \textbf{second} thermodynamic relation which has to be satisfied is $s=(\partial p/\partial T)_{\mu}$. While $p$ is given by \eqref{T33_a_3}, $s$ can be found from the Bekenstein-Hawking formula. The latter gives the entropy density of the boundary theory as the following
\begin{equation}
s=\frac{1}{V_3}\frac{A_3}{4G_5}=\frac{1}{4G_5}\frac{\int dx_1dx_2dx_3 \,e^{2V(\overline{r}_{+})+W(\overline{r}_{+})}}{\int dx_1dx_2dx_3}
\end{equation} 
where $A_3$ is the area of the horizon and $V_3$ is the voulme of the system in the boundary theory.
Considering $2 V(r)+W(r)=3 \ln r+B^2 S_2(r)$ and by using \eqref{metric_corrections}, we obtain
\begin{equation}\label{s}
s=\frac{(\pi T)^3}{4G_5}\left(1+\frac{2\mu^2}{(\pi T)^2}\right)+\frac{B^2}{8\pi G_5 T}\left(1-\frac{\mu^2}{(\pi T)^2}(2+3 a_3(T,\mu)+8k^2 \ln 2)\right)
\end{equation}
Interestingly, by using \eqref{T33_a_3} and equating $s=(\partial p/\partial T)_{\mu}=(\partial T^{33}/\partial T)_{\mu}$ with $s$ from \eqref{s}, we find $a_0$ and $a_1$ as exactly as given by \eqref{a_0_a_1}. It might be regarded as the second consistency check for the result \eqref{a_0_a_1}.

Having trusted to \eqref{a_0_a_1}, we now put it back into the stress tensor and charge current components to get the physical quantities on the boundary. 
In addition, we use the information about $\mathcal{N}=4$ SYM theory \cite{Son:2009tf}
\begin{equation}\label{dic}
\frac{1}{8\pi G_5}=\frac{N_c^2}{4\pi^2},\,\,\,\,\,\,\,\,k=-\frac{2}{\sqrt{3}},\,\,\,\,\,\,\,c=-\frac{k}{2 \pi G_5}=\frac{2 N_c^2}{\pi^2 \sqrt{3}}
\end{equation}
and write down all expression in terms of boundary field theory data, namely $T$, $\mu$, $N_c$, $B$ and $\Delta$. 
\begin{align}\label{T00_a_3_final}
T_{00}=&\frac{N_c^2}{8\pi^2}\big(3(\pi T)^4+12 (\pi T)^2\mu^2+8 \mu^4\big)+\frac{N_c^2B^2}{4 \pi^2}\bigg((1- \ln \frac{\pi T}{\Delta})-\frac{2}{3}\frac{\mu^2}{\pi T^2}(8 \ln 2-3)\bigg)\\\label{T33_a_3_final}
T_{ii}=&\frac{N_c^2}{24\pi^2}\big(3(\pi T)^4+12 (\pi T)^2\mu^2+8 \mu^4\big)+\frac{N_c^2B^2}{4 \pi^2}\bigg( \ln \frac{\pi T}{\Delta}+\frac{2}{3}\frac{\mu^2}{\pi T^2}(8 \ln 2-3)\bigg)\\\label{J0_a_3_final}
J^0 =& \frac{N_c^2}{3\pi^2}\big(3(\pi T)^2\,\mu+4 \mu^3\big)+\frac{N_c^2B^2}{3\pi^2}\frac{\mu}{(\pi T)^2}\left(8 \ln2-3\right)
\end{align}
As usual, the appearance of the transcendental numbers, like $\ln 2$, might be the sign of non-perturbative results. The latter is indeed consistent with the fact that our computations in the gravity side are related to a gauge theory at strong coupling on the boundary.
Now everything is ready to evaluate the magneto-conductivities obtained in \eqref{magneto_transport_from_hydro} for the $SU(N_c)$ gauge theory under study in the current section.

\subsection{Holographic magneto-conductivities in the hydrodynamic limit}
Let us recall that the magneto-conductivities computed in \sec{magneto_transport_from_hydro} are related to a general quantum anomalous system with weakly broken symmetries. At the limit $\tau_m \rightarrow \infty$, $\tau_e \rightarrow \infty$ and $\tau_c\rightarrow\infty$, our results reduces to those of a system with non-broken symmetries.  In this subsection we focus on the latter case. The system to which we would like to apply the hydro results is the SYM $SU(N_c)$ gauge theory studied in previous subsections.. The only thing we need is to compute the thermodynamic derivatives
in \eqref{epsilon_thermo} and \eqref{n_thermo}. Using \eqref{T00_a_3_final} and \eqref{J0_a_3_final}, we first compute $f_1$, $f_2$, $e_1$ and $e_2$. Then the conductivities of subsection \ref{magneto_transport_from_hydro} simplify to
\begin{align}
\sigma =&\, \frac{i}{\omega}\frac{2N_c^2\,\mu^2}{\pi^2} \left(1 + \frac{2 B^2 }{3 \pi^2 T^2 \mu^2}\right) ,\nonumber\\
T\alpha =& \,\frac{i}{\omega} N_c^2 T^2\,\mu \left(1 -\frac{2\mu^2}{3 \pi^2 T^2} + \frac{B^2}{3 \pi^4 T^4} \left(8 \ln2 - 7\right)\right),\nonumber\\\label{final_result_s}
T\kappa =& \,\frac{i}{\omega}\frac{\pi^2 N_c^2\, T^4}{2}\left(1 + \frac{4\mu^4}{3 \pi^4
	T^4} + \frac{B^2}{2 \pi^4 T^4} \left(1 - \frac{8 \mu^2}{3\pi^2 T^2}  \left(8 \ln2 -5\right)\right)\right).
\
\end{align}
As one expects, the above conductivities obey the constraints of Ward identities, discussed in \sec{gettin_back}. Let us denote that the order of the truncation (in the $\mu/T$ expansion) in the three conductivities above is such that in each of the constraints (\eqref{ward_1} and \eqref{ward_2}), both sides are at the	 same order.    

An important feature of the above results is that just like in the non-magnetic part,  
the magnetic part of conductivities scale quadratically with $N_c$ in the large $N_c$ limit. This means that the ratio of them would be independent of $N_c$.

Interestingly, the Wiedemann-Franz law, namely the statement that $\kappa/T\sigma$ is equal to $\pi^2/3$ in a Fermi liquid, fails in our system. It would be expected because our system was studied in the high temperature limit $\mu\ll T$.

Our results might be compared with those recently obtained in
\cite{Li:2018ufq}. While the result of the mentioned paper have been obtained numerically in the whole range of magnetic field, our analytic hydrodynamic results are valid in the $B\ll \mu^2\ll T^2$ limit. From their plots, a rough quadratic behavior at this limit is observed which is in agreement with our result.
\subsection{Anomalous transport from magnetized brane}
Previous computations of anomalous transport coefficients, specifically the chiral magnetic effect\footnote{$\sigma_{B}$ and $\sigma_{B}^{\epsilon}$ in \eqref{tau_mu_nu} and \eqref{nu_mu}.} from gauge gravity duality have been done in the background of RN black hole either by reading the gradient of gauge field on the boundary \cite{Son:2009tf} or by directly reading the corresponding Kubo formulas \cite{Landsteiner:2012kd}. The perturbative solution found in \cite{DHoker:2009ixq} and developed in the current paper allows us to  read the anomalous coefficients from boundary thermodynamic stress tensor and charge current. This is physically due to the fact that a weak magnetic field, as taken in the current paper, is equivalent to a long wavelength gauge field on the boundary. Such gauge field is consistent with the derivative counting of hydrodynamics \cite{Son:2009tf}. 

Let us consider \eqref{T_03} and \eqref{A_r}. When using the transformations \eqref{trans_T} and \eqref{trans_J} with \eqref{a_0_a_1}, we may rewrite them as 
\begin{align}
J^3 &=  -\frac{1}{2\pi G_5} B \,k\, \mu=\,c\,\mu\,B\\
T^{03}&=-\frac{1}{4\pi G_5}\,B \,k \,\mu^2=\,\frac{c}{2}\,\mu^2\, B
\end{align}
where we have also used \eqref{dic}. From the above one reads
\begin{equation}\label{chiral_trans}
\sigma_{B}=c\,\mu,\,\,\,\,\,\,\,\,\,\,\sigma_{B}^{\epsilon}=\frac{c}{2}\,\mu^2
\end{equation}
in complete agreement with the results of  Jensen et al.\cite{Jensen:2012jh} and Landsteiner et al.\cite{Landsteiner:2012kd} in the laboratory frame  and also with that of Son-Surowka \cite{Son:2009tf} and Banerjee et al.\cite{Banerjee:2012iz} \footnote{We would like to thank the anonymous referee for denoting the point that anomalous transport coefficients have been computed from the equilibrium partition function as well.} in the Landau-Lifshitz frame \footnote{See \cite{Kovtun:2012rj,Abbasi:2017tea} for detailed discussion on the hydrodynamic frames.}.

\section{Conclusion and Outlook}
\label{conclusion}
In the following we firstly review the set up and the new results found in this paper.  Then we discuss on some follow-up directions.

By the idea of computing the magneto-conductivities in an anomalous system with weakly broken symmetries, we generalized the hydro model given in \cite{Landsteiner:2014vua} to the second order in derivatives. The weakly broken symmetries in this model are characterized by three relaxation time parameters $\tau_m$, $\tau_e$ and $\tau_c$, corresponding to breaking the spacial translation, time translations and a $U(1)$ global symmetry. The anomaly enters in the model via the anomalous transport coefficients together with the ABJ term in the right hand side of the charge (non-)conservation equation.
Then we assumed the fluid to be coupled to slowly varying background fields by turning on a weak electric field and a small thermal gradient. The dynamical equations of such fluid, in the linear regime, are given by \eqref{E1}, \eqref{E2} and \eqref{E3}. In the language of the recently developed theory of quasi-hydrodynamics \cite{Grozdanov:2018fic}, these equations  are the linearized equations of anomalous quasi-hydrodynamics. By finding a stationary solution for them, namely \eqref{delta_T}, \eqref{delta_mu} and \eqref{delta_u},  we obtained our central results in the paper;  i.e. the magneto-conductivities \eqref{sigma}, \eqref{alpha1}, \eqref{alpha2} and \eqref{kappa}. These formulas show that our model is not  in general Onsager reciprocal, namely $\alpha_1\ne\alpha_2$. On the other hand, all of the conductivities depend on both chiral anomaly and the mixed gauge-gravitational anomaly coefficient.

The  formulas of conductivities mentioned above are general in the sense that they can be applied to the hydrodynamic regime of every arbitrary system. To this end, we just have to know the thermodynamics of the system. Knowing that, we can compute the thermodynamic derivatives $e_1$, $e_2$, $f_1$ and $f_2$ introduced in \eqref{epsilon_thermo} and \eqref{n_thermo} and then evaluate the conductivity formulas. We have indeed done it for two different systems; one in the weak and the other in the strong coupling regime.

In the weak regime the system which we considered  was a relativistic Weyl fluid in the presence of a magnetic field. The thermodynamics of this system had been found in \cite{Abbasi:2018zoc} before. We used the thermodynamic results of the mentioned paper and computed the conductivities at the limit $\mu\gg T$; see equations  \eqref{sigma_hydro} to \eqref{kappa_hydro}. The same conductivities had been found from the chiral kinetic theory in \cite{Abbasi:2018zoc} as well. However, in the latter case, the relaxation of the system had been characterized by just one relaxation time parameter in the kinetic equation, $\tau$. 
By comparing equations \eqref{sigma_hydro} to \eqref{kappa_hydro} with their counterparts in \cite{Abbasi:2018zoc}, we found that our general quasi-hydrodynamic model could be Onsager reciprocal if $\tau_m=\tau_e=\tau_c\equiv\tau$ \footnote{An Onsager reciprocal model for magneto transport in WSM was already constructed in \cite{Lucas:2016omy}.}. Interestingly we showed that at the same case, the conductivities obey the constraints found from Ward identities in \cite{Abbasi:2018zoc}. This simply suggests that in the limit $T \tau \gg1$, the equations of (anomalous) quasi-hydrodynamics can be derived from a generating functional. In other words there might be some generalized diff and gauge transformations which lead to these equations (See Appendix \ref{geneating}.).

In the strong regime, by motivations coming from both quark gluon plasma and gauge-gravity duality, we considered a $SU(N_c)$ SYM gauge theory at $N_c\gg1$ in the presence of a weak external magnetic field. As mentioned above, to evaluate the general hydro conductivity formulas  \eqref{sigma}, \eqref{alpha1}, \eqref{alpha2} and \eqref{kappa} in this system, we needed just to know the thermal state of the system. This state in the  dual gravity picture is described by magnetized charged brane. 
In the small magnetic field limit $B\ll T^2$ which corresponds to the hydro regime, the gravity solution has been found in  \cite{DHoker:2009ixq} formally in terms of some unevaluated integrals. By entering a second expansion, namely expansion over the $\mu/\pi T$, we could pertubatively evaluate all the metric and field strength components in the bulk up to an undetermined parameter, $a_3$ (See \eqref{metric_corrections}.).

To find the thermodynamics of the boundary gauge theory from the bulk solution, we encountered several problems.  First, when finding the stress tensor on the boundary,   a term $\ln \frac{\pi T}{L}$ including the radius of AdS5, $L$, appeared in our results. Since the boundary quantities have to be purely given in terms of boundary data, we entered  a counter term $\sim \ln r_0$ in \eqref{AdS_dictionary}. The radius $r_0$ in the bulk corresponds to an energy scale $\Delta$ in the gauge theory as $r_0=L^2 \Delta$.  Analogue of such cutoff emerging had been observed before in \cite{Abbasi:2018zoc}. In the mentioned paper it had been shown that in order to regulate the divergent integrals in the phase space, one would require to remove the states within a certain inner sphere in the Fermi sphere. This could be regarded in agreement with the basic requirement of chiral kinetic theory to exclude the purely quantum region from the phase space \cite{Stephanov:2012ki}.   
In our present case in the current paper, however, the origin of such energy scale (or cutoff) in the gauge theory, is not clear to us. We leave more investigation on it to a future work.

The second point we encountered with was displacing the horizon of the bulk solution due to the magnetic field. Finding the corrected horizon \eqref{corrected_r_H}, we recomputed the temperature $T$ and the chemical potential $\mu$.  Equations \eqref{trans_T} and \eqref{trans_J} simply say that the boundary temperature and chemical potential are not $T_0$ and $\mu_0$ which originally we started to work with them. In the following we replaced  $T_0$ and $\mu_0$ in terms of $T$ and $\mu$ everywhere in the computations.

The last problem with our gravity computations was the the parameter $a_3$.  To fix its value, we resorted the thermodynamic equalities on the boundary. We showed that \eqref{ansatz} with \eqref{a_0_a_1} leads to all the thermodynamic equalities be satisfied. 
 
 Having found the components of the stress tensor and charge current \eqref{T00_a_3_final}, \eqref{T33_a_3_final} and \eqref{J0_a_3_final}, we computed the thermodynamic derivatives $e_1$, $e_2$, $f_1$ and $f_2$ for SYM gauge theory. Using them, we then obtained our final results for magneto-conductivities in \eqref{final_result_s}.

Let us denote that the holographic thermoelectric transport has been studied in many papers through different models 		\cite{Davison:2016ngz,Donos:2014cya,Mokhtari:2017vyz}. However, the magnetized brane solution allows us to compute the anomalous transport coefficients as well.  As the last result, by using the pertutbative solution developed in the paper, we computed the chiral anomaly contribution to the chiral magnetic effect, \eqref{chiral_trans}, in agreement with the previous results in the literature.

It would be interesting to compute the effect of mixed gauge-gravitational anomaly on the magneto-conductivities. To this end one has to add a mixed gauge-gravitational coupling 
\begin{equation*}
\sim \epsilon^{MNJKL}A_{M}R^{A}_{\,\,BNJ}R^{B}_{\,\,AKL}
\end{equation*}
to the bulk action. The coefficient of this term contributes to the anomalous transport in a non-trivial way via the jump in the derivative expansion \cite{Landsteiner:2012kd,Jensen:2012kj}. We leave finding the gravity solution and computing the magneto-conductivities as well as anomalous coefficients in such set-up to a future work 		\cite{Abbasi:20191zoc}.  

What was studied in the current paper belongs to the family of quasi-hydrodynamics theories. Such recently developed hydrodynamic theories not only are interesting from the effective field theory view point 	\cite{Grozdanov:2018fic}, but also might be important phenomenologically  both in high energy physics 	\cite{Stephanov:2017ghc,Landsteiner:2014vua,Roychowdhury:2015jha,Sun:2016gpy} and condensed matter physics \cite{Gooth:2017mbd,Hartnoll:2007ih,Lucas:2017idv,Rogatko:2018lhn,Rogatko:2018moa}.

In this paper we studied magneto-transport in a chiral fluid with only one anomalous axial (weakly broken) symmetry. It would be interesting to compute the magneto-conductivities in a more realistic model with $U_V(1)\times U_A(1)$ charges.
\section*{Acknowledgment}
We would like to thank M. Chernodub, R. Davison,  D. Kharzeev and K. Landsteiner for useful discussions. We would also like to thank S. Ben-Abbas,  P. Janghorban, K. Kiaei, K. Naderi and P. Rezaei for discussion. 
We are grateful to Prof. F. Ardalan and Prof. H. Arfaei for their encouragements. A. Ghazi would like to thank M. M. Najafabadi for his support.
 A. Ghazi and O. Tavakol would like to thank school of particles and accelerators of IPM for hosting during this work.
\appendix

\section{Thermal and electric sources}
\label{Sources}
In the absence of external fields, the equilibrium solution of \eqref{EoMT} and \eqref{EoMJ} in our paper is given by
\begin{equation}\label{A_1}
\boldsymbol{E}=0:\,\,\,\,\,\,\,	u^{\mu}=(1,\boldsymbol{0}),\,\,\,\,\,T=T_0,\,\,\,\,\,\mu=\mu_0
\end{equation}
By turning on an external weak electric field, $\delta \boldsymbol{E}$, \eqref{A_1} is no longer the solution of equations, and so the fluid is driven from  equilibrium:
\begin{equation}
\underline{\delta \boldsymbol{E}}:\,\,\,\begin{cases}
u^{\mu}=(1,\boldsymbol{0})+\delta u^{\mu}(x)&\\
T=T_0+\delta T(x)&\\
\mu=\mu_0+\delta \mu(x)&
\end{cases} \quad
\end{equation}
where the $\delta$ contributions are sub-leading compared to the equilibrium values; \textbf{they are actually the linear response of the system to the external electric field $\delta \boldsymbol{E}$.}
\textbf{However}, there exists another way
of exciting the fluid;  heating a specific region of the fluid by  an external heater. This generates an external thermal gradient in the system, namely a slowly-varying temperature function $\text{T}(\boldsymbol{x})$. Again, this cannot be an equilibrium solution and the fluid is driven from equilibrium
\begin{equation}\label{3}
\,\,\,\begin{cases}
u^{\mu}=(1,\boldsymbol{0})+\delta u^{\mu}(x)&\\
T=\underline{\text{T}(\boldsymbol{x})}+\delta T(x)&\\
\mu=\mu_0+\delta \mu(x)&
\end{cases} \quad
\end{equation}
The $\delta$ terms are sub-leading compared to the first terms in the RHS's of the above equations;  \textbf{they are actually the linear response of the system to the background $\text{T}(\boldsymbol{x})$.}
Throughout the paper we have assumed the temperature has a gradient in the $z$-direction, $\text{T}(z)$. On the other hand, from eq. \eqref{magnet_conductivity} of the paper it is obvious that  we have to work with $-\nabla_z T/T$. Following \cite{Lucas:2016omy} and for sake of clarity, in the revised version of the paper we take the source as being
\begin{equation}\label{4}
\delta\zeta_i=-\frac{\nabla_i \text{T}}{\text{T}}=\,-\nabla_i \log \text{T}.
\end{equation}
In the equations of motion, we encounter with two types of temperature's derivatives: time derivative and $z-$derivative.
From \eqref{3}\footnote{Note that
	\begin{equation*}
	\log T=\log \text{T}+\log(1+\frac{\delta T}{\text{T}})=\log \text{T}+\frac{\delta T}{\text{T}}=\log \text{T}+\frac{\delta T}{T}+O(\delta^2)
	\end{equation*} from which one finds equation \eqref{A_6}.} and \eqref{4} above we may write:
\begin{eqnarray}
\partial_t T&=& 0+\,\partial_t \delta T(x)\\\label{A_6}
\frac{\nabla_z T}{T}& =&-\delta\zeta_z +\,\nabla_z (\frac{\delta T}{T})
\end{eqnarray}
As mentioned in the paper, we would like to work with the Fourier transformed fields in the limit $k\rightarrow 0$. We obtain:
\begin{eqnarray}
\partial_t T\,\,&\text{Fourier transformed}&=\,\,\,0+\,\-i \omega \delta \hat{T}\\
\nabla_z T\,\,\,\,&\text{Fourier transformed}&=\,\,\,\,- T_0\, \delta\hat{\zeta}_z+\,0
\end{eqnarray}
Let us denote that in  \eqref{deltaTtz}-\eqref{delta_j_z} in the paper we have not used the equations of motion yet, so the presence of $\zeta_z$ therein, i.e. in \eqref{deltaT_ii}, is due to the first order derivatives of the constitutive relations \eqref{tau_mu_nu} and \eqref{nu_mu}.

In summary, we see that in our set-up, the time-derivative of temperature appears as the perturbation $\delta\hat{T}$, while its $z-$derivative yields the source term $\delta\zeta_z$.
\section{Hydrodynamic linear response in an anomalous fluid with three relaxation time parameters}
\label{hydro_general_result}
As mentioned in the text, the fluid responds to the external sources
by small change in the hydrodynamic variables. When the relaxation in energy, momentum and charge are assumed to occur at different rates, say $\tau_e$, $\tau_m$ and $\tau_c$, respectively, one finds
\begin{align}
\delta\nonumber \hat{T}=&\frac{-i B}{w_0(e_2f_1-e_1f_2)}\left\{c\,\left(\frac{n\mu}{\omega_{m}}(f_1\,\mu-e_1)+\frac{}{}w(\frac{e_1}{\omega_{c}}-\frac{f_1\,\mu}{\omega_{e}})\right)+c_g\,\frac{ nT^2f_1}{\omega_m}\right\}\delta E_z^{(0)}\\ 
&+\frac{-i B}{w(e_2f_1-e_1f_2)}\left\{c\,\frac{\mu s}{\omega_{m}}\,(e_1-f_1\mu)+\frac{}{}c_g\,f_1T(\frac{w}{\omega_{e}}-\frac{sT}{\omega_{m}})\right\}\nabla_z T^{(0)}+\cdots
\end{align}
\begin{align}\nonumber 
\delta\hat{\mu}=&\frac{i B}{w(e_2f_1-e_1f_2)}\bigg\{c\,\left(\frac{n\mu}{\omega_{m}}(f_2\,\mu-e_2)+w(\frac{e_2}{\omega_{c}}-\frac{f_2\,\mu}{\omega_e})\right)+\frac{}{}c_g\, \frac{n T^2f_2}{\omega_m}\bigg\}\delta E_z^{(0)}\\
&
+\frac{B}{w(e_2f_1-e_1f_2)}\left\{c\,\frac{\mu s}{\omega_m}\,(e_2-f_2\mu)\,+\frac{}{}c_g\,f_2T_0\left(\frac{w }{\omega_{e}}- \frac{s T}{\omega_{m}}\right)\right\}\nabla_z T^{(0)}+\cdots\\\nonumber
\end{align}
\begin{align}\nonumber
\delta \hat{u_z}=&\,\frac{i}{\omega_{m}}\frac{n_0}{w_0}\,\delta E_z^{(0)}-\frac{i}{\omega_{M}}\frac{s_0}{w_0}\,\nabla_z T^{(0)}\,\\ \nonumber
&\,+\frac{i B^2}{w^2(e_2f_1-e_1f_2)}\biggl\{c^2\,\mu\left(\frac{\mu\, n}{\omega_m}(e_2-f_2\mu)+w(\frac{f_2\mu}{\omega_e}-\frac{e_2}{\omega_c})\right)\\\nonumber
&\,\,\,\,\,\,\,\,\,\,+ c_g^2\,\frac{f_1\,nT^3}{\omega_m}+c\,c_g\,T\left(w(\frac{e_1}{\omega_c}-\frac{f_1\mu_0}{\omega_e})-\frac{\mu n}{\omega_m}(e_1-f_1\mu+f_2T)\right)\biggr\}\delta E_z^{(0)}\\  \nonumber
&
\,+\frac{i B^2}{w^2(e_2f_1-e_1f_2)}\biggl\{c^2\,\frac{s\mu^2}{\omega_{m}}(f_2\mu-e_2)\,+c_g^2\,f_1T^2(\frac{w}{\omega_{e}}-\,\frac{sT}{\omega_{m}})\\
&\,\,\,\,\,\,\,\,\,\,\,\,\,\,\,\,\,\,\,\,\,\,\,\,\,\,\,\,+c\,c_g\,T\mu_0\left(-\frac{f_2w}{\omega_e}+\frac{s}{\omega_m}(e_1+f_2T-f_1\mu)\right)\biggr\}\nabla_z T^{(0)}+\cdots .
\end{align}

\section{Constraints from time reversal symmetry of microscopic dynamics in a system with weakly broken symmetries}
\label{App_Onsager}
Following \cite{Landau}, we define the entropy $S(x_i)$ in the system as the function of fluctuating quantities $x_i$ and write the probability of distribution as $w\, dx_1dx_2\cdots dx_n$ with the probability given as 
\begin{equation}
w=\,\text{constant} \,e^{S}:\,\,\,\,\,\,\,\,S=S_0-\frac{1}{2}\beta_{ij}\,x_ix_j
\end{equation}
where $S_0$ is the equilibrium entropy and $\beta_{ij}$ is symmetric ($\beta_{ij}=\beta_{ji}$) and positive-definite. Let us now define the thermodynamic conjugate of quantity $x_i$:
\begin{equation}\label{X}
X_i=\,-\frac{\partial S}{\partial x_i}=\,\beta_{ij}x_j
\end{equation}
One can immediately show that
\begin{equation}
\langle x_i x_j \rangle \sim \beta^{-1}_{ij},\,\,\,\,\,\langle x_i X_j \rangle \sim  \delta_{ij},\,\,\,\,\,\,\langle X_i X_j \rangle\sim \beta_{ij}.
\end{equation}
Close to equilibrium the fluctuating quantity $x_i$ may be given as the following
\begin{equation}
\dot{x}_i=-\lambda_{ij}x_j-\frac{x_i}{\tau_{(i)}}
\end{equation}
where the second term is indeed $\delta_{ij}x_{j}/\tau_{(j)}$ and therefore can be absorbed into the first one. However, we insist on writing the second term separately which allows us to simply restore the results regarding a system without relaxation by taking the limit $\tau_{(i)}\rightarrow \infty$. Considering  \eqref{X} one can now rewrite the above equation as 
\begin{equation}\label{Onsager_dynamics_modified}
\dot{x}_i=-\gamma_{ij}X_j-\frac{x_i}{\tau_{(i)}}
\end{equation}
where we have defined 
\begin{equation}
\gamma_{ij}=\lambda_{ik}\,\beta^{-1}_{kj}.
\end{equation}
On the other hand the time reversal symmetry of microscopic dynamics in equilibrium leads to
\begin{equation}
\langle x_i(t)\,x_j\rangle=\langle x_i\,x_j(t)\rangle.
\end{equation}
Differentiating with respect to the time and using \eqref{Onsager_dynamics_modified} we find
\begin{equation}
\gamma_{il}\,\langle X_l\,x_j\rangle+\frac{1}{\tau_{(i)}}\,\langle x_i\,x_j\rangle=\gamma_{jl}\,\langle x_i\,X_l\rangle+\frac{1}{\tau_{(j)}}\,\langle x_i\,x_j\rangle
\end{equation}
By considering the fact that $\langle x_i X_j\rangle\sim\, \delta_{ij}$, one obtains
\begin{equation}\label{T_constraint}
\gamma_{ij}+\,\frac{1}{\tau_{(i)}}\langle x_i\,x_j\rangle=\gamma_{ji}+\,\frac{1}{\tau_{(j)}}\langle x_i\,x_j\rangle
\end{equation}
This result is the constraint coming from the time reversal symmetry in a system with weakly broken symmetries. Let us reconsider equation \eqref{Onsager_dynamics_modified} as it follows
\begin{equation}
\left(\partial_t+\frac{i}{\tau_{(i)}}\right)x_i=-\gamma_{ij}X_j
\end{equation}
If one demands the kinetic coefficients $\gamma_{ij}$ in this equation are still reciprocal,   \eqref{T_constraint} forces all the relaxation parameters, $\tau_{(i)}$'s, to be the same.

\section{Quasi-hydrodynamics equations from generating functional }
\label{geneating}
Quasi-hydrodynamic equations given in \eqref{E1}, \eqref{E2} and \eqref{E3} can be phenomenologically used to describe the hydrodynamics in a metal in the presence of impurities \cite{Hartnoll:2007ih,Lucas:2017idv}. These phenomenological equations portray the non-conservation of energy momentum tensor $T^{\mu\nu}$ and $U(1)$ current $J^\mu$. In this subsection we would like to discuss how when the relaxation time parameters are of the order of macroscopic scale of variations in the system, or in other words when $\frac{1}{\tau}\sim \partial$, the quasi-hydrodynamic equations can be derived from a generating functional, however, with modified gauge and diff transformations.

To get the usual conservation equations the 
generating functional should satisfy
\begin{align}\label{eq-gen1}
W[A_\mu]&=W[\tilde{A}_\mu]\\
W[g_{\mu\nu}]&=W[\tilde{g}_{\mu\nu}]
\end{align}
 where
\begin{equation}
\tilde{g}_{ab}(\tilde{x})=g_{\mu\nu}({x})\frac{\partial{x^\mu}}{\partial{\tilde{x}^a}}\frac{\partial{x^\nu}}{\partial{\tilde{x}^b}}
\end{equation}
\begin{equation}
\tilde{A}_\mu=A_\mu +\partial_\mu \lambda
\end{equation}
Let us now consider the following transformations
\begin{align}
g_{ab}(x )&=\tilde{g}_{\mu\nu}(\tilde{{x}})(\frac{\partial{\tilde{x^\mu}}}{\partial{x^a}}+\frac{\tilde{u}_a}{\tau}(\tilde{x}^\mu-x^\mu))(\frac{\partial{\tilde{x^\nu}}}{\partial{x^b}}+\frac{\tilde{u}_b}{\tau}(\tilde{x}^\nu-x^\nu))\\
A_\mu&=\tilde{A}_\mu +(\partial_\mu+\frac{\tilde{u}_\mu}{\tau})\lambda
\end{align}
Applying a modified gauge transformation to the generating functional
\begin{equation}
\delta W =\int d^4 x \frac{\delta W}{\delta A_\mu} (\partial_\mu+\frac{\tilde{u}_\mu}{\tau})\lambda = \int d^4 x [-\partial_\mu J^\mu+\frac{\tilde{u}_\mu}{\tau}J^\mu]\lambda
\end{equation}
and demanding $\delta W =0$, we immediately find
\begin{equation}
\partial_\mu J^{\mu}= \frac{J^{\mu} \tilde{u}_\mu}{\tau}.
\end{equation}
Similarly, we can vary the generating functional by a modified diff transformation
\begin{align}\nonumber
\delta W &=\int d^4 x \frac{\delta W}{\delta g_{\mu \nu}} [\partial_\mu \xi_\nu+\frac{\tilde{u}_\mu}{\tau}\xi_\nu+\partial_\nu \xi_\mu+\frac{\tilde{u}_\nu}{\tau}\xi_\mu]\\
& = \int d^4x [-\partial_\mu T^{\mu\nu}+\frac{\tilde{u}_\mu T^{\mu\nu}}{\tau}]\xi_\nu+ [-\partial_\nu T^{\mu\nu}+\frac{\tilde{u}_\nu T^{\mu\nu}}{\tau}]\xi_\mu
\end{align}
The invariance of the generating functional under such  transformation then gives
\begin{align}
\partial_\mu T^{\mu0}&= \frac{T^{\mu 0} \tilde{u}_\mu}{\tau}\\
\partial_\mu T^{\mu i}&= \frac{T^{\mu i} \tilde{u}_\mu}{\tau}.
\end{align}
We leave more investigation on the topic to a future work \cite{Abbasi:2019zoc}.
\section{Extrinsic curvature}
\label{extrinsic}
The extrinsic curvature is defined as
$\mathcal{K}_{\mu \nu} = -\frac{1}{2} \left(\nabla_\mu n_\nu+ \nabla_\nu n_\mu\right) = \Gamma^\alpha_{\mu \nu} n_\alpha$
where $n_{\mu}$ is the unite outward normal vector to the boundary.
The non-vanishing components of the extrinsic curvature on the boundary are as the following
\begin{align}
\mathcal{K}_{00}&=- \frac{1}{2}\sqrt{U(r)} \left( 2 e^{2 W(r)} C(r)^2W'(r) + 2 e^{2 W(r)} C'(r)C(r)- U'(r)\right),\nonumber\\
\mathcal{K}_{30}&=\mathcal{K}_{03}= - \frac{1}{2}\sqrt{U(r)} e^{2W(r)} \left(C'(r)+ 2 C(r) W'(r)\right),\nonumber\\
\mathcal{K}_{11}&=\mathcal{K}_{22}= - e^{2V(r)} \sqrt{U(r)} V'(r),\nonumber\\
\mathcal{K}_{33}&= - e^{2W(r)} \sqrt{U(r)} W'(r)
\end{align}
and its trace is given by
\begin{equation}
\mathcal{K} = \gamma^{\mu \nu} \mathcal{K}_{\mu \nu} = -\frac{1}{U(r)}\mathcal{K}_{tt} + 2 \frac{C(r)}{U(r)}\mathcal{K}_{zt} + 2 e^{-2 V(r)}\mathcal{K}_{xx} -\left(\frac{C(r)^2}{U(r)} -e^{-2W(r)} \right)\mathcal{K}_{zz}.
\end{equation}
Considering $A=Q^2-3Mr^2$, the on-shell values read
 \begin{align}
\mathcal{K}_{00}=&\frac{\sqrt{U_{0}(r)}}{r^5}\,(r^4 U_0(r) -A-M r^2)\\\nonumber
&+\frac{B^2}{48\,r^7}\left(\sqrt{U_{0}(r)}\big(\frac{9k^2Q^4}{r_{+}^8}+8r^4\left(3a_3 + 4 \log r \right) \big)+\frac{4(A + M r^2 -r^4 U_0(r))}{\sqrt{U_{0}(r)}}(2 + 3 a_{3} + 4 \log r)\right)\\\nonumber
\mathcal{K}_{ii}=&-r\, \sqrt{U_0(r)}\\\nonumber
&+\frac{B^2}{270\,r^7\,\sqrt{U_0(r)}}\left(\frac{135}{2} r^6 a_3 + 90 r^4\left(U_0(r) \left(b + \log r\right) + r^2 \log r\right)-15A-\frac{8k^2Q^2}{r^2r_{+}^4}(A+3r^6)\right)\\\nonumber
\mathcal{K}_{33}=&-r \,\sqrt{U_{0}(r)}-\frac{B^2}{270\,r^7\,\sqrt{U_0(r)}}\frac{8k^2Q^2}{r^2r_{+}^4}(A+3r^6)\\\nonumber
&+\frac{B^2}{270\,r^7\,\sqrt{U_0(r)}}\left(\frac{135}{2} r^6 a_3 -30\left((A + 3 r^4 U_0(r))\log r + 6 r^4 U_0(r) b\right)+45r^4\left(r^2 + 2 U_0(r)\right)\right)\\
\mathcal{K}_{03}=&B\, k Q^2\,\frac{ \sqrt{U_0(r)}}{4r^3r_{+}^4}
\end{align}
\begin{align}
\mathcal{K}=&\frac{-r^2U_0(r) + M -3r^4}{r^3 \sqrt{U_0(r)}}\\\nonumber
&+\frac{B^2}{60 r^7 (U_0(r))^{\frac{3}{2}}} \left(5\left(3a_3 + 4 \log r +2\right) \left(Q^2 - 2 M r^2\right) + 4 U_0(r) \left( 5 r^4 - \frac{3 k^2 Q^2 U_0(r)}{r_+^4}\right)\right).
\end{align}

\bibliographystyle{utphys}

\begin{thebibliography}{10}
	
\bibitem{Fukushima:2008xe} 
K.~Fukushima, D.~E.~Kharzeev and H.~J.~Warringa,
``The Chiral Magnetic Effect,''
Phys.\ Rev.\ D {\bf 78}, 074033 (2008)
[arXiv:0808.3382 [hep-ph]].

	
	\bibitem{Son:2009tf}
	D.~T.~Son and P.~Surowka,
	``Hydrodynamics with Triangle Anomalies,''
	Phys.\ Rev.\ Lett.\  {\bf 103} (2009) 191601,
	[arXiv:0906.5044 [hep-th]].
	
	\bibitem{Landsteiner:2016led} 
	K.~Landsteiner,
	``Notes on Anomaly Induced Transport,''
	[arXiv:1610.04413[hep-th]].
	
		\bibitem{Landsteiner:2012kd} 
	K.~Landsteiner, E.~Megias and F.~Pena-Benitez,
	``Anomalous Transport from Kubo Formulae,''
	Lect.\ Notes Phys.  {\bf 871}, 433 (2013)
	[arXiv:1207.5808 [hep-th]].
		
	\bibitem{Banerjee:2008th} 
	N.~Banerjee, J.~Bhattacharya, S.~Bhattacharyya, S.~Dutta, R.~Loganayagam, and P.~Surowka,
	``Hydrodynamics from charged black branes,''
	JHEP {\bf 1101}, 094 (2011),
	[arXiv:0809.2596 [hep-th]].
	
	
	\bibitem{Erdmenger:2008rm} 
	J.~Erdmenger, M.~Haack, M.~Kaminski, and A.~Yarom,
	``Fluid dynamics of R-charged black holes,''
	JHEP {\bf 0901}, 055 (2009),
	[arXiv:0809.2488 [hep-th]].
	
	
	
		
	\bibitem{Landsteiner:2011cp} 
	K.~Landsteiner, E.~Megias and F.~Pena-Benitez,
	``Gravitational Anomaly and Transport,''
	Phys.\ Rev.\ Lett.\  {\bf 107}, 021601 (2011)
	[arXiv:1103.5006 [hep-ph]].
	

	
	
	\bibitem{Landsteiner:2011iq} 
	K.~Landsteiner, E.~Megias, L.~Melgar and F.~Pena-Benitez,
	``Holographic Gravitational Anomaly and Chiral Vortical Effect,''
	JHEP {\bf 1109}, 121 (2011)
	[arXiv:1107.0368 [hep-th]].
	
		
	\bibitem{Abbasi:2015saa} 
	N.~Abbasi, A.~Davody, K.~Hejazi and Z.~Rezaei,
	``Hydrodynamic Waves in an Anomalous Charged Fluid,''
	Phys.\ Lett.\ B {\bf 762}, 23 (2016),
	[arXiv:1509.08878 [hep-th]].


\bibitem{Yamamoto:2015ria} 
N.~Yamamoto,
``Chiral Alfvén Wave in Anomalous Hydrodynamics,''
Phys.\ Rev.\ Lett.\  {\bf 115}, no. 14, 141601 (2015)
[arXiv:1505.05444 [hep-th]].
	
	\bibitem{Chernodub:2015gxa} 
	M.~N.~Chernodub,
	``Chiral Heat Wave and mixing of Magnetic, Vortical and Heat waves in chiral media,''
	JHEP {\bf 1601}, 100 (2016)
	JHEP01(2016)100
	[arXiv:1509.01245 [hep-th]].
	
	\bibitem{Chernodub:2018era} 
	M.~N.~Chernodub, A.~Cortijo and K.~Landsteiner,
	``Zilch vortical effect,''
	Phys.\ Rev.\ D {\bf 98}, no. 6, 065016 (2018)
	[arXiv:1807.10705 [hep-th]].
	
		
	
	
	
	
	
	
	
	
	\bibitem{Kharzeev:2010gd}
	D.~E.~Kharzeev and H.~U.~Yee,
	``Chiral Magnetic Wave,''
	Phys.\ Rev.\ D {\bf 83} (2011) 085007,
	[arXiv:1012.6026  [hep-th]].
	
	
	\bibitem{Jensen:2012kj} 
	K.~Jensen, R.~Loganayagam and A.~Yarom,
	``Thermodynamics, gravitational anomalies and cones,''
	JHEP {\bf 1302}, 088 (2013)
	[arXiv:1207.5824 [hep-th]].
	
	
\bibitem{Nilson}
H. B. ~Nielsen and M. ~Ninomiya
``Adler-Bell-Jackiw Anomaly And Weyl Fermions In
Crystal,''
Phys. Lett. B 130, 389 (1983).



\bibitem{Weylsemimeta:Son}
D. T. ~Son, B. Z. ~Spivak
``Chiral Anomaly and Classical Negative Magnetoresistance of Weyl Metals,''
Phys .Rev. B 88 (2013) 104412 [arXiv:11206.1627  [cond-mat]]

\bibitem{Gorbar:2013dha} 
E.~V.~Gorbar, V.~A.~Miransky and I.~A.~Shovkovy,
``Chiral anomaly, dimensional reduction, and magnetoresistivity of Weyl and Dirac semimetals,''
Phys.\ Rev.\ B {\bf 89}, no. 8, 085126 (2014)
[arXiv:1312.0027 [cond-mat.mes-hall]].



\bibitem{Li:2017mbd} 
Q.~Li {\it et al.},
``Observation of the chiral magnetic effect in ZrTe5,''
Nature Physics 12, 550-554 (2016)
[arXiv:1412.6543 [cond-mat.str-el]].

		
	\bibitem{Gooth:2017mbd} 
	J.~Gooth {\it et al.},
	``Experimental signatures of the mixed axial-gravitational anomaly in the Weyl semimetal NbP,''
	Nature {\bf 547}, 324 (2017)
	[arXiv:1703.10682 [cond-mat.mtrl-sci]].
	
	\bibitem{Landsteiner:2014vua} 
	K.~Landsteiner, Y.~Liu and Y.~W.~Sun,
	``Negative magnetoresistivity in chiral fluids and holography,''
	JHEP {\bf 1503}, 127 (2015)
	[arXiv:1410.6399 [hep-th]].
	
	
	
	\bibitem{Grozdanov:2018fic} 
	S.~Grozdanov, A.~Lucas and N.~Poovuttikul,
	``Holography and hydrodynamics with weakly broken symmetries,''
	arXiv:1810.10016 [hep-th].
	
	
	\bibitem{Grozdanov:2017kyl} 
	S.~Grozdanov and N.~Poovuttikul,
	``Generalised global symmetries and magnetohydrodynamic waves in a strongly interacting holographic plasma,''
	arXiv:1707.04182 [hep-th].
	
		\bibitem{Lucas:2017idv} 
		A.~Lucas and K.~C.~Fong,
		``Hydrodynamics of electrons in graphene,''
		J.\ Phys.\ Condens.\ Matter {\bf 30}, 053001 (2018)
		[arXiv:1710.08425 [cond-mat.str-el]].
		
	\bibitem{Son:2012wh} 
	D.~T.~Son and N.~Yamamoto,
	`Berry Curvature, Triangle Anomalies, and the Chiral Magnetic Effect in Fermi Liquids,''
	Phys.\ Rev.\ Lett.\  {\bf 109}, 181602 (2012)
	[arXiv:1203.2697 [cond-mat.mes-hall]].
	
	
	\bibitem{Gao:2012ix} 
	J.~H.~Gao, Z.~T.~Liang, S.~Pu, Q.~Wang and X.~N.~Wang,
	``Chiral Anomaly and Local Polarization Effect from Quantum Kinetic Approach,''
	Phys.\ Rev.\ Lett.\  {\bf 109}, 232301 (2012)
	[arXiv:1203.0725 [hep-ph]].
	
	
	\bibitem{Son:2012zy} 
	D.~T.~Son and N.~Yamamoto,
	``Kinetic theory with Berry curvature from quantum field theories,''
	Phys.\ Rev.\ D {\bf 87}, no. 8, 085016 (2013)
	PhysRevD.87.085016
	[arXiv:1210.8158 [hep-th]].
	
	
	\bibitem{Hidaka:2016yjf} 
	Y.~Hidaka, S.~Pu and D.~L.~Yang,
	``Relativistic Chiral Kinetic Theory from Quantum Field Theories,''
	Phys.\ Rev.\ D {\bf 95}, no. 9, 091901 (2017)
	[arXiv:1612.04630 [hep-th]].
	
	
	\bibitem{Chen:2015gta} 
	J.~Y.~Chen, D.~T.~Son and M.~A.~Stephanov,
	``Collisions in Chiral Kinetic Theory,''
	Phys.\ Rev.\ Lett.\  {\bf 115}, no. 2, 021601 (2015)
	[arXiv:1502.06966 [hep-th]].
	
	\bibitem{Chen:2014cla} 
	J.~Y.~Chen, D.~T.~Son, M.~A.~Stephanov, H.~U.~Yee and Y.~Yin,
	``Lorentz Invariance in Chiral Kinetic Theory,''
	Phys.\ Rev.\ Lett.\  {\bf 113}, no. 18, 182302 (2014)
	PhysRevLett.113.182302
	[arXiv:1404.5963 [hep-th]].
	
	
	\bibitem{Stephanov:2012ki} 
	M.~A.~Stephanov and Y.~Yin,
	``Chiral Kinetic Theory,''
	Phys.\ Rev.\ Lett.\  {\bf 109}, 162001 (2012)
	PhysRevLett.109.162001
	[arXiv:1207.0747 [hep-th]].
	
	\bibitem{Abbasi:2017tea} 
	N.~Abbasi, F.~Taghinavaz and K.~Naderi,
	``Hydrodynamic Excitations from Chiral Kinetic Theory and the Hydrodynamic Frames,''
	JHEP {\bf 1803}, 191 (2018)
	[arXiv:1712.06175 [hep-th]].
	
	\bibitem{Abbasi:2018zoc} 
	N.~Abbasi, F.~Taghinavaz and O.~Tavakol,
	``Magneto-Transport in a Chiral Fluid from Kinetic Theory,''
	arXiv:1811.05532 [hep-th].
	
	\bibitem{Abbasi:2019zoc} 
	N.~Abbasi, A.~Ghazi,
	``Work in progress.''
	
\bibitem{Maldacena:1997re} 
J.~M.~Maldacena,
``The Large N limit of superconformal field theories and supergravity,''
Int.\ J.\ Theor.\ Phys.\  {\bf 38}, 1113 (1999)
[Adv.\ Theor.\ Math.\ Phys.\  {\bf 2}, 231 (1998)]
[hep-th/9711200].
	
	\bibitem{Witten:1998qj} 
	E.~Witten,
	``Anti-de Sitter space and holography,''
	Adv.\ Theor.\ Math.\ Phys.\  {\bf 2}, 253 (1998)
	[hep-th/9802150].


	
\bibitem{DHoker:2009ixq} 
E.~D'Hoker and P.~Kraus,
``Charged Magnetic Brane Solutions in AdS (5) and the fate of the third law of thermodynamics,''
JHEP {\bf 1003}, 095 (2010)
[arXiv:0911.4518 [hep-th]].



	
	\bibitem{Kharzeev:2011ds}
	D.~E.~Kharzeev and H.~U.~Yee,
	``Anomalies and time reversal invariance in relativistic hydrodynamics: the second order and higher dimensional formulations,''
	Phys.\ Rev.\ D {\bf 84} (2011) 045025,
	[arXiv:1105.6360 [hep-th]].
	
	\bibitem{Grozdanov:2016tdf} 
	S.~Grozdanov, D.~M.~Hofman and N.~Iqbal,
	``Generalized global symmetries and dissipative magnetohydrodynamics,''
	Phys.\ Rev.\ D {\bf 95}, no. 9, 096003 (2017)
	[arXiv:1610.07392 [hep-th]].
	
	
	\bibitem{Hernandez:2017mch} 
	J.~Hernandez and P.~Kovtun,
	``Relativistic magnetohydrodynamics,''
	JHEP {\bf 1705}, 001 (2017)
	[arXiv:1703.08757 [hep-th]].
	

	\bibitem{Hartnoll:2007ih} 
	S.~A.~Hartnoll, P.~K.~Kovtun, M.~Muller and S.~Sachdev,
	``Theory of the Nernst effect near quantum phase transitions in condensed matter, and in dyonic black holes,''
	Phys.\ Rev.\ B {\bf 76}, 144502 (2007)
	[arXiv:0706.3215 [cond-mat.str-el]].
	
	\bibitem{Moore:2012tc} 
	G.~D.~Moore and K.~A.~Sohrabi,
	``Thermodynamical second-order hydrodynamic coefficients,''
	JHEP {\bf 1211}, 148 (2012)
	[arXiv:1210.3340 [hep-ph]].
	
	
	
	\bibitem{Hartnoll:2009sz} 
	S.~A.~Hartnoll,
	``Lectures on holographic methods for condensed matter physics,''
	Class.\ Quant.\ Grav.\  {\bf 26}, 224002 (2009)
	[arXiv:0903.3246 [hep-th]].
	
	
	\bibitem{Herzog:2009xv} 
	C.~P.~Herzog,
	``Lectures on Holographic Superfluidity and Superconductivity,''
	J.\ Phys.\ A {\bf 42}, 343001 (2009)
	[arXiv:0904.1975 [hep-th]].
		
		

	
	\bibitem{Hartnoll:2016apf}
	S.~A.~Hartnoll, A.~Lucas and S.~Sachdev,
	``Holographic quantum matter,''
	arXiv:1612.07324 [hep-th].
	
	
	
	\bibitem{Shuryak:2003xe} 
	E.~Shuryak,
	``Why does the quark gluon plasma at RHIC behave as a nearly ideal fluid?,''
	Prog.\ Part.\ Nucl.\ Phys.\  {\bf 53}, 273 (2004)
	doi:10.1016/j.ppnp.2004.02.025
	[hep-ph/0312227].
	
	\bibitem{Maldacena:1997re} 
	J.~M.~Maldacena,
	``The Large N limit of superconformal field theories and supergravity,''
	Int.\ J.\ Theor.\ Phys.\  {\bf 38}, 1113 (1999)
	[Adv.\ Theor.\ Math.\ Phys.\  {\bf 2}, 231 (1998)]
	[hep-th/9711200].
	
	\bibitem{Witten:1998qj} 
	E.~Witten,
	``Anti-de Sitter space and holography,''
	Adv.\ Theor.\ Math.\ Phys.\  {\bf 2}, 253 (1998)
	[hep-th/9802150].
	
	
	\bibitem{Balasubramanian:1999re} 
	V.~Balasubramanian and P.~Kraus,
	``A Stress tensor for Anti-de Sitter gravity,''
	Commun.\ Math.\ Phys.\  {\bf 208}, 413 (1999)
	[hep-th/9902121].
	
		
		
	\bibitem{Li:2018ufq} 
	W.~Li, S.~Lin and J.~Mei,
	``On Conductivities of Magnetic Quark-Gluon Plasma at Strong Coupling,''
	arXiv:1809.02178 [hep-th].
	
	\bibitem{Lucas:2016omy} 
	A.~Lucas, R.~A.~Davison and S.~Sachdev,
	``Hydrodynamic theory of thermoelectric transport and negative magnetoresistance in Weyl semimetals,''
	Proc.\ Nat.\ Acad.\ Sci.\  {\bf 113}, 9463 (2016)
	[arXiv:1604.08598 [cond-mat.str-el]].
	
	
	
	
	
	
	\bibitem{Davison:2016ngz} 
	R.~A.~Davison, W.~Fu, A.~Georges, Y.~Gu, K.~Jensen and S.~Sachdev,
	``Thermoelectric transport in disordered metals without quasiparticles: The Sachdev-Ye-Kitaev models and holography,''
	Phys.\ Rev.\ B {\bf 95}, no. 15, 155131 (2017)
	[arXiv:1612.00849 [cond-mat.str-el]].
	
	\bibitem{Donos:2014cya} 
	A.~Donos and J.~P.~Gauntlett,
	``Thermoelectric DC conductivities from black hole horizons,''
	JHEP {\bf 1411}, 081 (2014)
	[arXiv:1406.4742 [hep-th]].
	Kovtun:2012rj
	
	\bibitem{Mokhtari:2017vyz} 
	A.~Mokhtari, S.~A.~Hosseini Mansoori and K.~Bitaghsir Fadafan,
	``Diffusivities bounds in the presence of Weyl corrections,''
	Phys.\ Lett.\ B {\bf 785}, 591 (2018)
	[arXiv:1710.03738 [hep-th]].
	
	
	
	\bibitem{Abbasi:20191zoc} 
	N.~Abbasi, F.~Taghinavaz and O.~Tavakol,
	``Work in progress.''
	
	
	\bibitem{Stephanov:2017ghc} 
	M.~Stephanov and Y.~Yin,
	``Hydrodynamics with parametric slowing down and fluctuations near the critical point,''
	Phys.\ Rev.\ D {\bf 98}, no. 3, 036006 (2018)
	[arXiv:1712.10305 [nucl-th]].
	
	
	te{Roychowdhury:2015jha}
	\bibitem{Roychowdhury:2015jha} 
	D.~Roychowdhury,
	``Magnetoconductivity in chiral Lifshitz hydrodynamics,''
	JHEP {\bf 1509}, 145 (2015)
	[arXiv:1508.02002 [hep-th]].
	

	\bibitem{Sun:2016gpy} 
	Y.~W.~Sun and Q.~Yang,
	``Negative magnetoresistivity in holography,''
	JHEP {\bf 1609}, 122 (2016)
	[arXiv:1603.02624 [hep-th]].
	
	\bibitem{Rogatko:2018lhn} 
	M.~Rogatko and K.~I.~Wysokinski,
	``Magnetotransport of Weyl semimetals with $\mathbb{Z}_2$ symmetry and chiral anomaly,''
	arXiv:1810.07521 [hep-th].
	
	\bibitem{Rogatko:2018moa} 
	M.~Rogatko and K.~I.~Wysokinski,
	``Hydrodynamics of topological Dirac semi-metals with chiral and $\mathbb{Z}_2$ anomalies,''
	JHEP {\bf 1809}, 136 (2018)
	[arXiv:1804.02202 [hep-th]].
	
	\bibitem{Jensen:2012jh} 
	K.~Jensen, M.~Kaminski, P.~Kovtun, R.~Meyer, A.~Ritz and A.~Yarom,
	``Towards hydrodynamics without an entropy current,''
	Phys.\ Rev.\ Lett.\  {\bf 109}, 101601 (2012)
	[arXiv:1203.3556 [hep-th]].
	
	\bibitem{Banerjee:2012iz} 
	N.~Banerjee, J.~Bhattacharya, S.~Bhattacharyya, S.~Jain, S.~Minwalla and T.~Sharma,
	``Constraints on Fluid Dynamics from Equilibrium Partition Functions,''
	JHEP {\bf 1209}, 046 (2012)
	[arXiv:1203.3544 [hep-th]].
	
	\bibitem{Kovtun:2012rj} 
P.~Kovtun,
``Lectures on hydrodynamic fluctuations in relativistic theories,''
J.\ Phys.\ A {\bf 45}, 473001 (2012),
[arXiv:1205.5040[hep-th]].

	
			
		\bibitem{Landau} 
		L. ~Landau and E.~ Lifshitz,
``	Statistical Physics: Volume 5.''
			
	
	



	
	
	
	
	
	
	
	
	
	
	
	
	
	
		
	
	
	
	
	
	
	
	
	
	
	
	
	
	
	
	
	
	
	
	
	
	
	
	
	
	
	
	

	
	
	



	
	
	
	
	
	
	

	
	
	
	
	
	
	
	
	
	
	
	
	
	
	
	
	
	
	
	
	
	
	
	
	
	
	
	
	
	
	
	
	
	
	
	
	
	
	
	
	
	
	
	
	
\end{thebibliography}
\providecommand{\href}[2]{#2}\begingroup\raggedright\endgroup

\end{document}